\documentclass[aps,prd,onecolumn,noshowpacs,preprintnumbers,eqsecnum,nofootinbib]{revtex4} \usepackage[reqno]{amsmath}
\usepackage{epsfig}
\usepackage{array}
\usepackage{float}
\usepackage{lscape,graphicx}
\usepackage{amssymb}
\usepackage[usenames,dvipsnames]{color}
\usepackage{ulem}
\usepackage{enumerate}
\usepackage{natbib}

\def\ltap{\ \raisebox{-.4ex}{\rlap{$\sim$}} \raisebox{.4ex}{$<$}\ }
\def\gtap{\ \raisebox{-.4ex}{\rlap{$\sim$}} \raisebox{.4ex}{$>$}\ }

\newcommand{\betabeta}{\mbox{$(\beta \beta)_{0 \nu}  $}}

\newcommand{\meff}{\mbox{$\left|  < \!  m \!  > \right| $}}

\newcommand{\bea}{\begin{equation}\begin{array}{c}}
\newcommand{\eea}{\end{array}\end{equation}}
\newcommand{\ea}{\end{array}} 

\newcommand{\beq}{\begin{equation}}
\newcommand{\eeq}{\end{equation}}
\newcommand{\bad}{\begin{array}{ccc}}
\newcommand{\dmsol}{\mbox{$\Delta m^2_{\odot}$}}
\newcommand{\dma}{\mbox{$\Delta m^2_{\rm A}$ }}
\newcommand{\mefff}{\mbox{$ < \! m \! > $}}
\newcommand{\ba}{\begin{array}{c}}
\newcommand{\half}{\frac{1}{2}}
\newcommand{\diag}{{\rm diag}}


\begin{document}

\preprint{{\it SISSA 10/2011/EP}}
\preprint{{\it TUM-HEP 799/11}}
\preprint{{\it IPMU11--0043}}
\preprint{{\it CFTP/11-006}}

\title{Low Energy Signatures of the TeV Scale See-Saw Mechanism}

\author{A. Ibarra$^{a)}$, E. Molinaro$^{b)}$~and~S. T. Petcov$^{c,d,e)}$}

\affiliation{{\it $^{a)}$Physik-Department T30d, Technische Universit\"at M\"unchen, James-Franck-Stra{\ss}e, 85748 Garching, Germany}}
\affiliation{{\it $^{b)}$Centro de F\'{i}sica Te\'{o}rica de Part\'{i}culas (CFTP), Instituto Superior T\'{e}cnico,  Avenida Rovisco Pais 1, 
1049-001, Lisboa, Portugal}}
\affiliation{{\it $^{c)}$SISSA and INFN-Sezione di Trieste, Via Bonomea 265, 34136 Trieste, Italy}}
\affiliation{{\it $^{d)}$IPMU, University of Tokyo, Tokyo, Japan}}
\affiliation{{\it $^{e)}$Institute of Nuclear Research and Nuclear Energy, Bulgarian Academy of Sciences, 1784 Sofia, Bulgaria.}}

\begin{abstract}
We study a type I see-saw scenario 
where the right-handed (RH) neutrinos, responsible 
for the light neutrino mass generation, lie
at the electroweak scale. Under certain conditions, the strength of 
the charged (CC) and neutral current (NC) weak 
interactions of the Standard Model particles with the 
heavy RH neutrinos can be large enough 
to allow the production of the latter at the LHC, 
opening also the possibility of
observing other low energy signatures of the new physics in the electroweak
precision observables as well as in searches for rare leptonic decays 
or neutrinoless double beta decay.
In this scenario the flavour structure of the 
indicated CC and NC couplings of the heavy RH neutrinos 
is essentially determined by the low energy 
neutrino parameters, 
leading to fairly strong correlations among the
new phenomena.
In particular, we show that the present
bound on the $\mu\rightarrow e +\gamma$ decay rate 
makes very difficult the observation of the 
heavy RH neutrinos at the LHC or the observation of deviations
from the Standard Model predictions in the electroweak precision data.
 We also show that all present experimental 
constraints on this scenario
still allow i) for an enhancement of the rate of 
neutrinoless double beta decay, 
which thus can be in the
range of sensitivity
of the GERDA experiment even when 
the light Majorana neutrinos possess
a normal hierarchical mass spectrum,
and ii) for the  predicted 
$\mu\to e+ \gamma$ decay rate 
to be within the sensitivity range of the 
MEG experiment.

\end{abstract}

\maketitle

\section{Introduction}

One of the best motivated extensions of the Standard Model consists
of introducing extra fermions that are singlets under the Standard Model 
gauge group, which we will call for definiteness right-handed (RH) neutrinos.
With this particle content, the Lagrangian contains
a Yukawa coupling of the RH neutrinos
with the left-handed leptonic doublets and the Higgs doublet, which 
leads to a Dirac neutrino mass when 
the electroweak symmetry is spontaneously
broken. Besides, it contains a Majorana mass term 
for the RH neutrinos, which is a priori unrelated 
to the electroweak symmetry breaking scale. The most 
popular scenario in this model consists on assuming that 
the RH neutrino Majorana mass scale is much larger than 
the electroweak symmetry breaking scale, thus
naturally leading to tiny  Majorana neutrino masses through the renowned see-saw
mechanism~\cite{seesaw}. Furthermore, any other low energy effect of the 
RH neutrinos decouples at least with their mass squared, 
resulting in a  tiny rate for the rare leptonic decays~\cite{meg_SM}, a tiny 
leptonic electric dipole moment~\cite{Archambault:2004td} 
and a tiny deviation of the electroweak observables from the 
Standard Model predictions~\cite{EW,Antusch:2008tz,Antusch:2006vwa}, 
in excellent agreement with presently existing experimental data.

It should be borne in mind, however, that the mass of the 
RH neutrinos is a free parameter which can take any value
between zero and the Planck scale. 
An interesting possibility arises when the RH neutrinos have
masses in the range ${\cal O}(100\div1000)$ GeV. If this is the case,
the new particles could be produced and detected  at the 
Large Hadron Collider (LHC), if their couplings to the 
Standard Model particles are sizable~\cite{colliders,delAguila:2008hw}. 
This situation, at first sight bizarre in view of
the tininess of the neutrino masses, can be realized in some well
motivated scenarios, namely when lepton number is approximately 
conserved (see, e.g.,~\cite{GonzalezGarcia:1988rw,Kersten:2007vk,Gavela08}).
More importantly, the contributions from the new particles to
the low energy phenomena are no longer suppressed, opening the possibility
of finding additional low energy signatures of the new physics with
experiments at the intensity frontier. Furthermore, the existence of 
RH neutrinos with masses in the range of $(100 \div 1000)$ GeV 
with lepton number violating interactions can dramatically
enhance the rate of neutrinoless double beta decay~\cite{HPR83,JV83},
inducing rates that could possibly be observable
at GERDA~\cite{Abt:2004yk}, even when the light neutrinos present
a normal hierarchical mass spectrum.

In this paper we analyze the constraints from various
experiments on the scenario where the RH 
neutrinos are accessible to collider searches,
as well as the interrelation between the different constraints.
More specifically, we will derive the constraints that follow 
from the present bounds on the rate of the process 
$\mu\rightarrow e+\gamma$, with  a special emphasis
on the relation to the neutrino mixing and oscillation 
parameters. We will also discuss the prospects
to observe neutrinoless double beta decay in the next round of
experiments, in the view of all present 
experimental constraints.  In section
2 we discuss the formalism and define the parameter space
of the theory. In sections 3 and 4 we discuss the constraints on the parameter space
which arise from radiative charged lepton decays and the implications for
collider searches and electroweak precision observables. In the subsequent section
we perform a detailed analysis of the possible enhancement of the neutrinoless double beta decay 
rate due to the exchange of the heavy (RH) Majorana neutrinos. In section 6 we combine
the constraints on the parameter space that we obtained in sections 3, 4 and 5.
Finally, we report our conclusions.

\section{Preliminary Remarks}

We denote the light and heavy Majorana neutrinos with definite masses 
as $\chi_i$ and $N_k$, respectively. 
\footnote{We use the same notations as in \cite{Ibarra:2010xw}.}
The  charged and neutral current weak 
interactions involving  the light Majorana neutrinos have the form:
%%%%%%%%%%%%%%%%%%%%%%%%%%%%%%%%%%%%%%%%%%%%%
\begin{eqnarray}
\label{nuCC}
\mathcal{L}_{CC}^\nu 
&=& -\,\frac{g}{\sqrt{2}}\, 
\bar{\ell}\,\gamma_{\alpha}\,\nu_{\ell L}\,W^{\alpha}\;
+\; {\rm h.c.}
=\, -\,\frac{g}{\sqrt{2}}\, 
\bar{\ell}\,\gamma_{\alpha}\,
\left( (1+\eta)U \right)_{\ell i}\,\chi_{i L}\,W^{\alpha}\;
+\; {\rm h.c.}\,,\\
\label{nuNC} 
\mathcal{L}_{NC}^\nu &=& -\, \frac{g}{2 c_{w}}\,
\overline{\nu_{\ell L}}\,\gamma_{\alpha}\,
\nu_{\ell L}\, Z^{\alpha}\; 
= -\,\frac{g}{2 c_{w}}\,
\overline{\chi_{i L}}\,\gamma_{\alpha}\,
\left (U^\dagger(1+\eta+\eta^\dagger)U\right)_{ij}\,\chi_{j L}\,
Z^{\alpha}\,,
\end{eqnarray}
%%%%%%%%%%%%%%%%%%%%%%%%%%%%%%%%%%%%%%%%%%
%
where $(1+\eta)U = U_{\rm PMNS}$ is the 
Pontecorvo, Maki, Nakagawa, Sakata (PMNS)
neutrino mixing matrix \cite{BPont57,MNS62,BPont67}, 
$U$ is a $3\times 3$ unitary 
matrix which diagonalises the Majorana mass matrix 
of the left-handed (LH) flavour neutrinos $\nu_{\ell L}$
(generated by the see-saw mechanism), 
and the matrix $\eta$ characterises the deviations 
from unitarity of the PMNS matrix.
The elements of  $U_{\rm PMNS}$ are determined 
in experiments studying the oscillations of the 
flavour neutrinos and antineutrinos, $\nu_\ell$ and $\bar{\nu}_\ell$, 
$\ell=e,\mu,\tau$, at relatively low energies.
In these experiments the initial states of the 
flavour neutrinos, produced in some weak process,
are coherent superpositions of the states of 
the light massive Majorana neutrino $\chi_i$ only.
The states of the heavy Majorana neutrino $N_j$ are 
not present in the superpositions representing the 
initial flavour neutrino states
and this leads to deviations 
from unitarity of the PMNS matrix.

  The charged current and the neutral current  
weak interaction couplings of the heavy 
Majorana neutrinos $N_j$ to the $W^\pm$ and 
$Z^0$ bosons read:
%%%%%%%%%%%%%%%%%%%%%%%%%%%%%%%%%%%%%%%%%%%
\begin{eqnarray}
 \mathcal{L}_{CC}^N &=& -\,\frac{g}{2\sqrt{2}}\, 
\bar{\ell}\,\gamma_{\alpha}\,(RV)_{\ell k}(1 - \gamma_5)\,N_{k}\,W^{\alpha}\;
+\; {\rm h.c.}\,
\label{NCC},\\
 \mathcal{L}_{NC}^N &=& -\frac{g}{4 c_{w}}\,
\overline{\nu_{\ell L}}\,\gamma_{\alpha}\,(RV)_{\ell k}\,(1 - \gamma_5)\,N_{k}\,Z^{\alpha}\;
+\; {\rm h.c.}\,.
\label{NNC}
\end{eqnarray}
%%%%%%%%%%%%%%%%%%%%%%%%%%%%%%%%%%%%%%
%
Here $V$ is the unitary matrix which diagonalises the 
Majorana mass matrix of the heavy RH neutrinos and 
the matrix $R$ is determined by (see \cite{Ibarra:2010xw})
$R^* \cong M_D\, M^{-1}_{N}$, $M_D$ and $M_N$ being the 
neutrino Dirac and the RH neutrino Majorana mass matrices, 
respectively, $|M_D| \ll |M_N|$. The matrix $\eta$ which
parametrises the deviations from unitarity of the
neutrino mixing matrix, 
can be expressed in terms of the matrix $R$: 
%%%%%%%%%%%%%%%%%%%%%%%%%%%%%%%%%%%%
\begin{equation}
	\eta\;\equiv\;-\half R R^\dagger =  
-\half (RV)(RV)^\dagger = \eta^\dagger\,.
\label{eta}
\end{equation}
%%%%%%%%%%%%%%%%%%%%%%%%%%%%%%%%%%%%
%
It is possible to constrain 
the elements of the hermitian matrix $\eta$
by using the existing neutrino oscillation data and
data on electroweak (EW) processes 
\cite{Antusch:2008tz,Antusch:2006vwa} 
(e.g., on $W^{\pm}$ decays, invisible $Z$ decays, 
universality tests of EW interactions). For 
$M_k \gtap 100$ GeV these constraints read 
\cite{Antusch:2008tz,Antusch:2006vwa}:
%%%%%%%%%%%%%%%%%%%%%%%%%%%%%%%%%%%%%%%%%%%%%%
\begin{equation}
|\eta|\;<\;\left(
      \begin{array}{ccc}
       2.0\times 10^{-3} & 0.6\times 10^{-4} & 1.6\times 10^{-3}\\
       0.6\times 10^{-4} & 0.8\times 10^{-3} & 1.0\times 10^{-3}\\
       1.6\times 10^{-3} & 1.0\times 10^{-3} & 2.6\times 10^{-3}
      \end{array}\right)\,.\label{eta_bounds}
\end{equation}
%%%%%%%%%%%%%%%%%%%%%%%%%%%%%%%%%%%%%%%%%%%

 The elements of the matrix RV and the masses $M_k$ of the heavy 
Majorana neutrinos $N_k$ should satisfy the approximate constraint 
on the elements of the Majorana mass matrix of the LH flavour 
neutrinos \cite{Merle:2006du}, 
$|(m_{\nu})_{\ell'\ell}| \lesssim 1$ eV, $\ell,\ell'=e,\mu,\tau$.
In the case of the type I see-saw mechanism under discussion this implies:
%%%%%%%%%%%%%%%%%%%%%%%%%%%%%%%%%%%%%%%%%%%
\begin{equation}
\sum_{k} |(RV)^*_{\ell'k}\;M_k\, (RV)^{\dagger}_{k\ell}| \lesssim 1~{\rm eV}\,,
~\ell',\ell=e,\mu,\tau\,.
\label{VR1}
\end{equation}
%%%%%%%%%%%%%%%%%%%%%%%%%%%%%%%%%%%%%%%%%%
%
This relation can be satisfied in several situations.
The most trivial way to satisfy it is by imposing that
$|(RV)_{\ell k}|\ll 1$ for all $\ell$ and $k$,
which renders the observation of all RH neutrinos
impossible at the LHC or in any low energy phenomena. 
However, this relation can also be fulfilled if one element
of the matrix $RV$ is sizable. This requires the existence
of at least another large matrix element, in order to cancel
the large contribution of the former one in eq.~(\ref{VR1}).
Whereas the possibility of cancellations between three different
terms cannot be precluded, in the simplest case only two terms
will cancel. Then, the large contribution to eq.~(\ref{VR1}) from one of
the RH neutrinos, say $N_1$ with mass $M_1$,  
is cancelled by a negative contribution from another
RH neutrino, say $N_2$ with mass $M_2$, provided
%%%%%%%%%%%%%%%%%%%%%%%%%%%%%%%%%%
\begin{equation}
(RV)_{\ell 2}=\pm i\, (RV)_{\ell 1}\sqrt{\frac{M_1}{M_2}}
\label{rel0}
\end{equation}
%%%%%%%%%%%%%%%%%%%%%%%%%%%%%%%%%%
%
which is naturally fulfilled if the RH
neutrinos $N_1$ and $N_2$ form a pseudo-Dirac pair \cite{LW81,STPPD82},
i.e., if there exists an approximately 
conserved lepton charge (see, e.g., \cite{Ibarra:2010xw}).
In this scenario, in order not to spoil 
the cancellation between these two
terms, the contribution from the third neutrino to eq.~(\ref{VR1}) 
should be negligible. 
Therefore in what follows we will work for simplicity in the 
$3\times 2$ see-saw scenario,
in which the indicated CC and NC weak interaction couplings of 
$N_3$ are set to zero and $N_3$ is decoupled. 
In this case the SM is effectively 
extended by the addition of 2 RH neutrino fields only. 
In this class of models 
(see, e.g., \cite{3X2Models,Ibarra:2003up,PRST05})
one of the three light (Majorana) neutrinos
is massless and hence the neutrino mass spectrum is 
hierarchical. Two possible 
types of hierarchical spectrum are 
allowed by the current neutrino data 
(see, e.g., \cite{PDG10}):  
$i)$ normal hierarchical (NH),
$m_{1}=0$, $m_{2}=\sqrt{\dmsol}$ and $m_{3}=\sqrt{\dma}$, 
where $\dmsol \equiv m^2_{2} -  m^2_{1} > 0$ and
$\dma \equiv m^2_{3} -  m^2_{1}$;\\
$ii)$ inverted hierarchical (IH), 
$m_{3}=0$, $m_{2}=\sqrt{|\dma|}$ and 
$m_{1}=\sqrt{|\dma| -\dmsol} \cong \sqrt{|\dma|}$, 
where $\dmsol \equiv m^2_{2} -  m^2_{1} > 0$ and 
$\dma =  m^2_{3} -  m^2_{2} < 0$.
In both cases we have: $\dmsol/|\dma| \cong 0.03 \ll 1$.

  The two heavy neutrino fields satisfy the 
Majorana condition:
$C \overline{N_{k}}^T = N_k$, $k=1,2$. 
If the Majorana mass matrix $M_N$ of 
the RH neutrinos is not CP invariant, 
one can always make the eigenvalues of 
$M_N$ real and positive, $M_{1,2} > 0$.
For $M_N$ respecting the CP symmetry,
the two real eigenvalues of $M_N$ 
can have the same or opposite
signs (see, e.g., \cite{BiPet87}). 
With the choice $M_2 > 0$ 
we can always make (without 
loss of generality),  $M_1$ can be positive 
or negative: $M_1 > 0$ and 
$M_1 < 0$. One can show, however, 
that we get the same results for 
the observables of interest  
to this study in the two cases.
Therefore in what follows we 
shall work with $M_{1,2} > 0$.

     We assume that the heavy RH neutrinos $N_{1,2}$ have masses
in the range $M_{1,2}=\mathcal{O}(100\div1000)$ GeV,
which makes possible, in principle,
their production, e.g., at LHC.  
In order to be produced with observable rates at LHC,
the CC and NC couplings of $N_{1}$ and $N_2$ in 
eqs. (\ref{NCC}) and (\ref{NNC}) have to be sufficiently
large. Under this condition
the existing experimental upper bounds 
on the neutrinoless double beta (\betabeta) 
decay rate put stringent constraints on
the mass spectrum of the RH neutrinos. 
It can be shown \cite{Ibarra:2010xw}, in particular,
that in the type I see-saw scenario of interest 
the two heavy (RH) Majorana neutrinos $N_{1,2}$ 
must be almost degenerate in mass:
$M_2 \cong M_1$.
If we assume that $M_{2}> M_{1}>0$ and 
$M_{2}\equiv (1+z) M_{1}$, $z>0$,  
in order to satisfy the experimental limit on 
the \betabeta-decay rate, one should 
have $z\lesssim 10^{-3}$ ($10^{-2}$) for 
$M_{1}\approx 10^{2}$ 
($10^{3}$) GeV \cite{Ibarra:2010xw}.

  The charged current and the neutral current  
weak interaction couplings of the heavy 
Majorana neutrinos $N_j$,  $(RV)_{\ell k}$,
are furthermore constrained by the requirement
of reproducing the correct low energy 
neutrino oscillation parameters
after the decoupling of the heavy degrees of freedom.
Remarkably, under the condition that $|(RV)_{\ell k}|$ are 
sufficiently large  to produce observable effects of the RH
neutrinos at low energies, these couplings take a very simple 
form \cite{Ibarra:2010xw}.

Indeed, the Dirac mass matrix $M_{D}$ in the case under study can be 
written as \cite{CIbar01}
%%%%%%%%%%%%%%%%
\begin{equation}
M_{D}\;=\;i \,U_{PMNS}^{*}\sqrt{\hat{m}}\, O\, \sqrt{\hat{M}}V^{\dagger}\,,
\label{MDfromO}
\end{equation}
%%%%%%%%%%%%%%%%%
%
where $\hat{m}\equiv\diag(m_{1},m_{2},m_{3})$ 
and $O$ is a complex orthogonal matrix. In the 
scheme with two heavy RH Majorana neutrinos 
the matrix $O$ has the form~\cite{Ibarra:2003up}:
%%%%%%%%%%%%%%%%%%%%%%%%%%%%%%%%
\begin{eqnarray}
	O\;\equiv\;\left(\begin{array}{cc}
				0 & 0\\
				\cos\hat{\theta} & \pm\sin\hat{\theta}\\
				-\sin\hat{\theta} & \pm\cos\hat{\theta}
\end{array}\right)\,,
\;\;\;\;\;\;\;{\rm for\;NH\;mass\;spectrum}\\
	O\;\equiv\;\left(\begin{array}{cc}
				\cos\hat{\theta} & \pm\sin\hat{\theta}\\
				-\sin\hat{\theta} & \pm\cos\hat{\theta}\\
				0 & 0
\end{array}\right)\,,\;\;\;\;\;\;\;{\rm for\;IH\;mass\;spectrum}
\end{eqnarray}
%%%%%%%%%%%%%%%%%%%%%%%%%%%%%%%%
%
where $\hat{\theta}=\omega-i\xi$.
The RH neutrino mixing matrix entering into the CC and NC 
weak interaction Lagrangians (\ref{NCC}) and (\ref{NNC}) can be 
expressed as~\cite{Ibarra:2010xw}
%%%%%%%%%%%%%%%%%%%%%%%%%%%%%%%%
\begin{equation}
	RV \;=\; 
-i \,U_{PMNS}\, \sqrt{\hat{m}}\, O^{*}\, \sqrt{\hat{M}^{-1}}\,.
\end{equation}
%%%%%%%%%%%%%%%%%%%%%%%%%%%%%%%%
%
The $O$-matrix in the case of, e.g., NH spectrum can be 
decomposed as follows:
%%%%%%%%%%%%%%%%%%%%%%%%%%%%%%%%
\begin{eqnarray}
	O\;=\;
\frac{e^{i\hat\theta}}{2}
\left(
\begin{array}{cc}
0 & 0\\
1 & \mp i\\
i & \pm 1
\end{array}
\right) + 
\frac{e^{-i\hat\theta}}{2}
\left(
\begin{array}{cc}
0 & 0\\
1 & \pm i\\
-i & \pm 1
\end{array}
\right) 
= O_{+} + O_{-}\,.
\label{Opm}
\end{eqnarray}
%%%%%%%%%%%%%%%%%%%%%%%%%%%%%%%%%%
%
One can get a similar expression for the IH spectrum.
The Dirac neutrino mass matrix can be decomposed accordingly
as $M_D=M_{D+}+M_{D-}$, in a self-explanatory notation.
Taking for definiteness $\xi>0$, it follows that $M_{D+}$ ($M_{D-}$) 
grows (decreases) exponentially with
$\xi$.~\footnote{ Obviously, if $\xi < 0$, $M_{D-}$ ($M_{D+}$) 
will grow (decrease) exponentially with $\xi$. 
It is possible to show that for sufficiently large 
values of $|\xi|$ of interest,
the results for the different observables 
considered in our study do not depend 
on the choice of the sign of $\xi$.}~Therefore, for sufficiently
large  $\xi$ it is possible to compensate the huge suppression
in eq.~(\ref{MDfromO}) from the tiny observed neutrino masses 
and the relatively light RH neutrino masses.~\footnote{Note, however, that $M_{D-}$
cannot be neglected in the calculation of the 
Majorana mass matrix of the LH flavour neutrinos 
even though it is exponentially suppressed 
compared to $M_{D+}$: the naive approximation 
$M_D\simeq M_{D+}$ leads to $m_\nu= 0$, due to $O_{+}O_{+}^T=0$ 
(see \cite{Ibarra:2010xw} for details).
Therefore, reproducing the correct 
light neutrino masses and mixing
requires a large amount of fine tuning,
unless the RH neutrinos form
a pseudo-Dirac pair.
Demanding $(M_{D})_{ij}\sim{\cal O}(1\,{\rm GeV})$
and  random RH neutrino masses of ${\cal O}(100\,{\rm GeV})$,
for instance, 
requires a tuning of one part in $10^9$ in order 
to produce a neutrino mass 
$m_i \sim{\cal O}(10^{-2}\,{\rm eV})$.}

We are interested in heavy Majorana neutrino couplings 
to charged leptons and gauge bosons, which are large enough to
produce observable low energy signatures.
In the limit of ``large'' $\xi$, 
the matrix $O$ in eq. (\ref{Opm})
can be very well approximated by:
%%%%%%%%%%%%%%%%%%%%%%%%%%%%%%%%%%%%%%%%%%%%%%%%%%%%
\begin{equation}
	O\;\approx\; \frac{e^{i\omega}e^{\xi}}{2}\,\left(\begin{array}{cc}
				0 & 0\\
				1 & \mp i\\
				i & \pm 1
			\end{array}\right)\,.
\end{equation}
%%%%%%%%%%%%%%%%%%%%%%%%%%%%%%%%%%%%%%%%%%
%
For NH spectrum and ``large'' $\xi$, 
the matrix $RV$ takes the form:
%%%%%%%%%%%%%%%%%%%%%%%%%%%%%%%%
\begin{equation}
RV\;\approx\;- \frac{e^{-i\omega} e^{\xi}}{2}\,
\sqrt{\frac{m_{3}}{|M_{1}|}}\left(
      \begin{array}{cc}
\left(U_{e3}+i\sqrt{m_{2}/m_{3}}\,U_{e2}\right) &   \pm i\left(U_{e3}+i\sqrt{m_{2}/m_{3}}\,U_{e2}\right)/\sqrt{1+z} \\
\left(  U_{\mu 3}+i\sqrt{m_{2}/m_{3}}\,U_{\mu 2}\right ) & \pm i\left(U_{\mu 3}+i\sqrt{m_{2}/m_{3}}\,U_{\mu 2}\right)/\sqrt{1+z} \\
\left( U_{\tau 3}+i\sqrt{m_{2}/m_{3}}\,U_{\tau 2}\right ) & \pm i\left(U_{\tau 3}+i\sqrt{m_{2}/m_{3}}\,U_{\tau 2}\right)/\sqrt{1+z}   
      \end{array}\right)\,,
\label{RVpseudo}
\end{equation}
%%%%%%%%%%%%%%%%%%%%%%%%%%%%%%%%
In a different context similar expressions for the corresponding neutrino Yukawa couplings were derived in 
\cite{Strumia04} in a scheme in which a successful leptogenesis can take place at $T\sim 10^7$ GeV, and in
\cite{Gavela08} where a TeV scale see-saw model with approximately conserved lepton charge was proposed.
In the case of IH light neutrino mass spectrum, 
the matrix $RV$ is obtained by replacing $m_{2,3}\to m_{1,2}$ and  
$U_{\alpha 2, \alpha 3}\to U_{\alpha 1,\alpha 2}$ ($\alpha=e,\mu,\tau$)  
in eq.~(\ref{RVpseudo}). 
For both types of neutrino mass spectrum, the elements of 
the two columns of $RV$ in the limit of ``large'' $\xi$ 
are related by the condition: 
%%%%%%%%%%%%%%%%%%%%%%%%%%%%%%%%%%
\begin{equation}
(RV)_{\alpha 2}=\pm i\, (RV)_{\alpha 1}/\sqrt{1+z}
\,\;\;\;\;\;{\rm for}\;\;\alpha=e,\mu,\tau.
\label{rel1}
\end{equation}
%%%%%%%%%%%%%%%%%%%%%%%%%%%%%%%%%%
%
thus recovering eq.~(\ref{rel0}).

If $M_N$ is CP conserving 
and $M_1 < 0$, $M_2 >0$, the 
expressions for $(RV)_{\alpha 1}$ 
in eq. (\ref{RVpseudo})
have an additional factor $-i$ and 
instead of eq.~(\ref{rel1}) we have:
$(RV)_{\alpha 2}=\mp \, (RV)_{\alpha 1}/\sqrt{1+z}$.
This relation and the relation given 
in eq.~(\ref{rel1}) lead to the same expressions 
for the observables discussed 
further in the present study.

The overall size of the couplings eq.~(\ref{RVpseudo}) depends
crucially on the value of the parameter $\xi$
which has 
 no direct
 physical interpretation. 
 It proves convenient to express $\xi$ in terms of the 
largest eigenvalue $y$ of the matrix of neutrino Yukawa couplings
using the relation:
%%%%%%%%%%%%%%%%%%%%%%%%%%%%%%%%%
\begin{equation}\label{maxy}
y^{2}v^{2}\;\equiv\;\text{max}\left\{\text{eig}\left(M_{D}M_{D}^{\dagger}\right)
\right\}
\;=\; 
\text{max}\left\{
\text{eig}\left(\sqrt{\hat{m}}O\hat{M}O^{\dagger}\sqrt{m}\right)
\right\}
\;=\;
\frac{1}{4}e^{2\xi}M_{1}(m_{2}+m_{3})(2+z)\,,
\end{equation}
%%%%%%%%%%%%%%%%%%%%%%%%%%%%%%%%%
%
with $v=174$ GeV. In terms of $y$ and 
for $z\ll1$, the heavy Majorana neutrino couplings become:
%%%%%%%%%%%%%%%%%%%%%%%%%%%%%%%
\begin{eqnarray}
\label{mixing-vs-y}
\left|\left(RV\right)_{\alpha 1} \right|^{2}&=&
\frac{1}{2}\frac{y^{2} v^{2}}{M_{1}^{2}}\frac{m_{3}}{m_{2}+m_{3}}
	\left|U_{\alpha 3}+i\sqrt{m_{2}/m_{3}}U_{\alpha 2} \right|^{2}\,,
~~{\rm NH}\,,\\
\left|\left(RV\right)_{\alpha 1} \right|^{2}&=&
\frac{1}{2}\frac{y^{2} v^{2}}{M_{1}^{2}}\frac{m_{2}}{m_{1}+m_{2}}
	\left|U_{\alpha 2}+i\sqrt{m_{1}/m_{2}}U_{\alpha 1} \right|^{2}
\cong \;\frac{1}{4}\frac{y^{2} v^{2}}{M_{1}^{2}}
\left|U_{\alpha 2}+iU_{\alpha 1} \right|^{2}\,,
\,{\rm IH}\,,
\label{mixing-vs-yIH}
\end{eqnarray}
%%%%%%%%%%%%%%%%%%%%%%%%%%%%
%
where we have used the fact that for the IH spectrum one has
$m_1 \cong m_2$.

From eqs.~(\ref{mixing-vs-y}) or (\ref{mixing-vs-yIH}), using the unitarity
of the matrix $U$, one can express the neutrino Yukawa eigenvalue $y$ in (\ref{maxy}) in terms
of heavy neutrino to charged leptons coupling constants:
\begin{equation}\label{ymax2}
	y^{2}v^{2}\;=\;2\,M_{1}^{2}\,\left(\left| (RV)_{e1} \right|^{2}+\left| (RV)_{\mu1} \right|^{2}+\left| (RV)_{\tau1} \right|^{2}\right)\,.
\end{equation} 
An upper limit on the neutrino Yukawa coupling $y$ can be derived by assuming 
the validity of perturbative unitarity. Indeed, the requirement of
perturbative unitarity can be easily fulfilled if the see-saw parameter space
satisfies the condition:
\begin{equation}\label{pertuni}
	\frac{\Gamma_{N_{i}}}{M_{i}}\;<\;\frac{1}{2}\,,
\end{equation}
where $\Gamma_{N_{i}}$ is the total decay rate of the heavy Majorana neutrino $N_{i}$, which in the limit
$M_{N_{i}}\gg v$ is given by the following expression:
\begin{equation}\label{gammaNi}
	\Gamma_{N_{i}}\;=\;\frac{g^{2}}{16\pi M_{W}^{2}}\, M_{i}^{3}\,\sum_{\ell}\left| (RV)_{\ell i} \right|^{2}\,,
\end{equation}
Therefore, taking into account eqs.~(\ref{ymax2}) and (\ref{gammaNi}) and the condition~(\ref{pertuni}), 
we get the following upper limit on the parameter $y$ from perturbative unitarity:
\begin{equation}
	y\;<\;4\,.
\end{equation}

%%%%%%%%%%%%%%%%%%%%%%%%%%%%%%%%%%%
%
\mathversion{bold}
\section{The $\mu \to e + \gamma$ Decay}
\mathversion{normal}
%
%%%%%%%%%%%%%%%%%%%%%%%%%%%%%%%%%%%

The $\mu\to e+\gamma$ decay branching ratio in the
scenario under discussion is given by 
\cite{Petcov:1976ff,Cheng:1980tp}:
%%%%%%%%%%%%%%%%%%%%%%%%%%%%%%%%%%
\begin{eqnarray}
B(\mu\to e+\gamma) =
\frac{\Gamma(\mu\to e+\gamma)}{\Gamma(\mu\to e+\nu_{\mu}+\overline{\nu}_{e})} 
&=& 
\frac{3\alpha_{\rm em}}{32\pi}\,|T|^{2}\,,
\label{Bmutoeg1}
\end{eqnarray}
%%%%%%%%%%%%%%%%%%%%%%
%
where $\alpha_{\rm em}$ is the fine structure constant and
%%%%%%%%%%%%%%%%%%%%%%%%%%%%%
\begin{eqnarray}
T&=&\sum\limits_{j=1}^{3}
\left[\left(1+\eta\right)U\right]_{\mu j}^{*} \,\left[\left(1+\eta\right)U \right]_{e j} G\left(\frac{m_{j}^{2}}{M_{W}^{2}}\right)\nonumber\\
&+& \sum\limits_{k=1}^{2} \left( RV\right)_{\mu k}^{*} \left( RV\right)_{e k} G\left(\frac{M_{k}^{2}}{M_{W}^{2}}\right)\,.
\end{eqnarray}
%%%%%%%%%%%%%%%%%%%%%%%%%%%%%%%%
%
The loop integration function $G(x)$ has the form:
%%%%%%%%%%%%%%%%%%%%%%%%%%%%%%%%%%%%%%%%
\begin{equation}
G(x)\;=\;\frac{10-43x+78 x^2 - 49 x^3 + 4 x^4 + 18 x^3 \log(x)}{3 (x - 1)^4}\,.
\end{equation}
%%%%%%%%%%%%%%%%%%%%%%%%%%%%%
%
It is easy to verify that $G(x)$ is a monotonic 
function which takes values in the interval $[4/3,10/3]$, with
$G(x)\cong\frac{10}{3}-x$ for $x\ll 1$. 
From the definition of the matrix $\eta$, 
eq.~(\ref{eta}), 
one can write the amplitude $T$ as follows:
%%%%%%%%%%%%%%%%%%%%%%%%%%%%
\begin{equation}
	T\;\cong\; \,
\left [(RV)_{\mu 1}^{*}\, (RV)_{e1}\,
+ (RV)_{\mu 2}^{*}\, (RV)_{e2}\right ] \left[ G(X) - G(0)\right]\,,
\label{T2}
\end{equation}
%%%%%%%%%%%%%%%%%%%%%%%%%%%
%
where $X \equiv(M_{1}/M_{W})^{2}$ and we have 
assumed that the difference between $M_1$ and $M_2$
is negligibly small and used $M_{1}\cong M_{2}$.
Using eqs.~(\ref{T2}) and (\ref{rel1}) we get:
%%%%%%%%%%%%%%%%%%%%%%%%%%%%
\begin{equation}
|T|\;\cong\; \,
\frac{2 + z}{1 + z}\,
\left |(RV)_{\mu 1}^{*}\, (RV)_{e1}\right | \left| G(X) - G(0)\right|\,.
\label{T3b}
\end{equation}
%%%%%%%%%%%%%%%%%%%%%%%%%%%
%
Finally, using the expressions of $|(RV)_{\mu1}|^{2}$
and  $|(RV)_{e 1}|^{2}$ in terms of neutrino parameters, 
eqs.~(\ref{mixing-vs-y}) and (\ref{mixing-vs-yIH}), we obtain 
the $\mu\to e+\gamma$ decay branching ratio for the NH and IH spectra:
%%%%%%%%%%%%%%%%%%%%%%%%%%%%
\begin{eqnarray}
&&{\rm\bf NH:}\;\;\;B(\mu\to e+\gamma)\cong\nonumber \\
&& \frac{3\alpha_{\rm em}}{32\pi}
\left(\frac{y^{2}v^{2}}{M_{1}^{2}}\frac{m_{3}}{m_{2}+m_{3}}\right)^{2}
\left|U_{\mu 3}+i\sqrt{\frac{m_{2}}{m_{3}}}U_{\mu 2} \right|^{2}
\left|U_{e 3}+i\sqrt{\frac{m_{2}}{m_{3}}}U_{e 2} \right|^{2}
\left[G(X)-G(0) \right]^{2}\,,
\label{meg-U-NH} \\\nonumber&&\\
&&{\rm\bf IH:}\;\;\;B(\mu\to e+\gamma)\cong\nonumber\\ 
&& \frac{3\alpha_{\rm em}}{32\pi}
\left(\frac{y^{2}v^{2}}{M_{1}^{2}}\frac{1}{2}\right)^{2}
\left|U_{\mu 2}+iU_{\mu 1} \right|^{2}
\left|U_{e 2}+iU_{e 1} \right|^{2}
\left[G(X)-G(0) \right]^{2}\,.
\label{meg-U-IH}
\end{eqnarray}
%%%%%%%%%%%%%%%%%%%%%%%%%%%
%
 Employing the standard parametrisation of 
the neutrino mixing matrix \cite{PDG10} it is 
not difficult to obtain expressions for the factors 
$|U_{\ell 3}+i\sqrt{m_{2}/m_{3}}U_{\ell 2}|^{2}$ 
and $|U_{\ell 2}+iU_{\ell 1}|^{2}$, $\ell=e,\mu$, 
in terms of the neutrino mixing parameters 
and the solar and atmospheric neutrino mass 
squared differences:
%%%%%%%%%%%%%%%%%%%%%%%%%%%%%%%%%%%%%%%%%%%
\begin{eqnarray}
&& 
\left|U_{e 3}+i\sqrt{\frac{m_{2}}{m_{3}}}U_{e 2} \right|^{2} =
s^2_{13}  + 
\left(\frac{\dmsol}{\dma}\right)^{1/2}
c^2_{13}\,s^2_{12} \nonumber \\
&&
\hspace{1.4cm}
 -\, 2\left(\frac{\dmsol}{\dma}\right)^{1/4}
c_{13}\,s_{13}\,s_{12}
\sin\left(\delta+\frac{\alpha_{21}-\alpha_{31}}{2}\right)\,, 
\label{Ue32NH}\\
&& 
\left|U_{\mu 3}+i\sqrt{\frac{m_{2}}{m_{3}}}U_{\mu 2} \right|^{2}=
c^2_{13}\,s^2_{23} + 
\left(\frac{\dmsol}{\dma}\right)^{1/2}\,
\left (c^2_{12}\,c^2_{23} + s^2_{12}\,s^2_{23} s^2_{13} - 
2 c_{12}\,c_{23}\,s_{12}\,s_{23} s_{13}\,\cos\delta \right)\nonumber \\
&& 
\hspace{0.4cm}
+\,
2 \left(\frac{\dmsol}{\dma}\right)^{1/4}
c_{13}\,s_{23}
\left [c_{12}c_{23}\sin\left(\frac{\alpha_{31}-\alpha_{21}}{2}\right) - 
s_{12}\,s_{23}\,s_{13}\,\sin\left(\frac{\alpha_{31}-\alpha_{21}}{2}-\delta\right) 
\right ]\,, \label{Umu32NH}\\
&& 
\left| U_{e 2} + i U_{e 1} \right|^{2} 
= c^2_{13}\left [ 1 
+ 2 c_{12}\,s_{12}\sin\left(\frac{\alpha_{21}}{2}\right)\right ]\,,
\label{Ue21IH}
\\
&& \left|U_{\mu 2}+i U_{\mu 1} \right|^{2} = 
c_{23}^2+s_{13}^2\,s_{23}^2 - 2\,c_{12}\,s_{12}\,(c_{23}^2-s_{13}^2\,s_{23}^2)
\sin\frac{\alpha_{21}}{2}\nonumber\\&&
\hspace{1.4cm}+\,2\,c_{23}\,s_{13}\,s_{23}
\left[s_{12}^2\sin\left(\frac{\alpha_{21}}{2}+\delta\right) - 
c_{12}^2\sin\left(\frac{\alpha_{21}}{2}-\delta\right) \right]\,,
\label{Umu21IH}
\end{eqnarray}
%%%%%%%%%%%%%%%%%%%%%%%%%%%%%%%%%%%%%%%%
% 
where $c_{ij} \equiv \cos\theta_{ij}$,  $s_{ij} \equiv \sin\theta_{ij}$,
$\theta_{12}$, $\theta_{23}$ and $\theta_{13}$ are 
respectively the solar neutrino, atmospheric neutrino 
and the CHOOZ angles, 
$\delta$ is the Dirac CP violating phase and 
$\alpha_{21}$ and $\alpha_{31}$ are the 
two Majorana CP violating phases \cite{BHP80}.

\begin{figure}
\begin{center}
\begin{tabular}{cc}
\includegraphics[width=7.5cm,height=6.5cm]{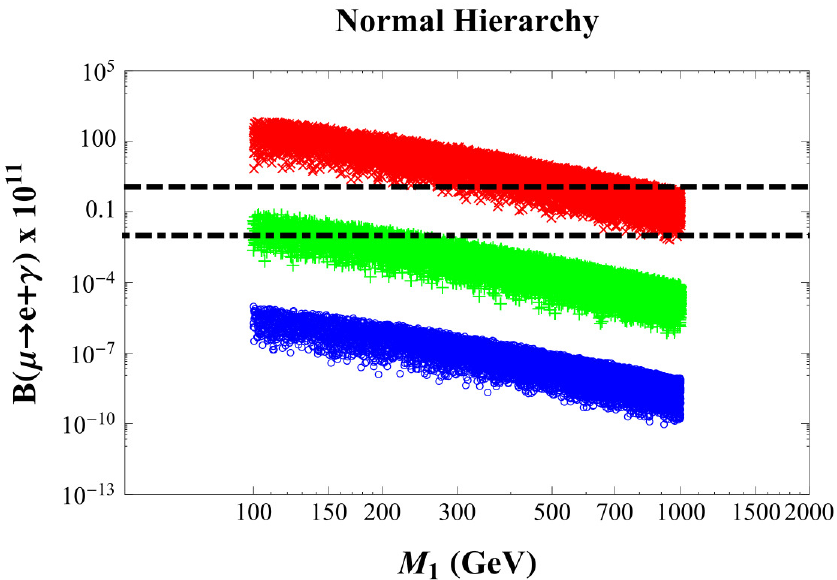} &
\includegraphics[width=7.5cm,height=6.5cm]{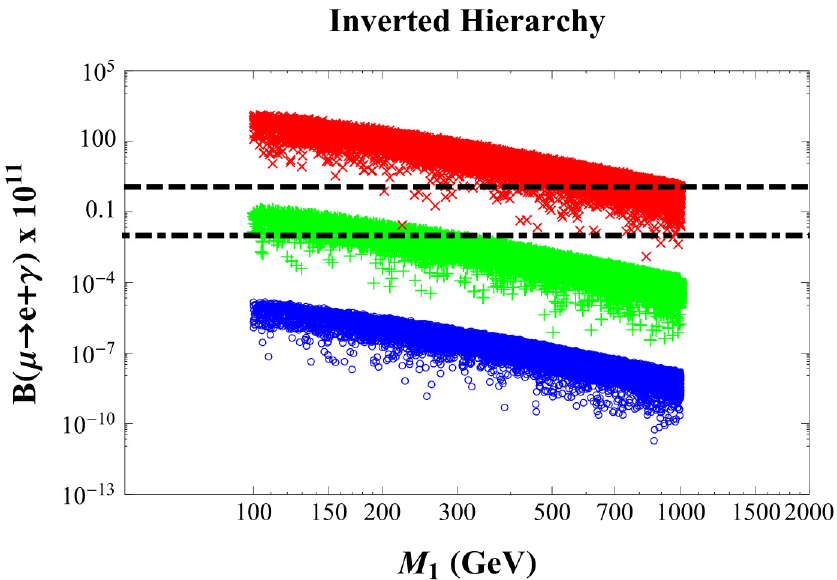}
\end{tabular}
\caption{The dependence of $B(\mu\to e+\gamma)$ on $M_{1}$ 
in the case of NH (left panel)
and IH (right panel) light neutrino mass spectrum, 
for $i)$ $y=0.001$ (blue $\circ$), 
$ii)$ $y=0.01$ (green $+$), and 
$iii)$ $y=0.1$ (red $\times$). 
The horizontal dashed line corresponds to the MEGA
bound \cite{MEGA}, $B(\mu\rightarrow e +\gamma)\leq
1.2 \times 10^{-11}$. The horizontal dot-dashed line
corresponds to $B(\mu\rightarrow e +\gamma) =
10^{-13}$, which is the prospective sensitivity
of the MEG experiment \cite{MEG}. 
\label{fig1}}
\end{center}
\end{figure}

  We show in Fig.~\ref{fig1} the branching ratio of
$\mu\to e+\gamma$ as a function of the 
RH neutrino mass $M_{1}$, 
for three different values 
of the neutrino Yukawa coupling eigenvalue: $y=0.001\, (0.01)\,[0.1]$, 
blue $\circ$ (green $+$) [red $\times$]. 
The figure was obtained by performing a scan 
of the values of the neutrino mixing angles
$\theta_{12}$, $\theta_{23}$ and $\theta_{13}$
and  the solar and atmospheric neutrino mass squared 
differences 
$\dmsol$ and $\dma$
within the corresponding 3$\sigma$ bounds 
(see Table~\ref{tab:mixing2010}).~\footnote{In all scatter plots included in our paper
the neutrino observables are scanned assuming a 
gaussian distribution with the corresponding mean 
value and standard deviation reported in Table 1. 
All the other (unmeasurable) see-saw parameters are
selected with a flat distribution.}~The Majorana phases 
$\alpha_{21}$ and $(\alpha_{31} - \alpha_{21})$ 
are varied in the interval~\footnote{We note that the phases $\alpha_{21}$ and $\alpha_{31}$
enter into the expression for the neutrino mixing
matrix in the form $\exp(i\alpha_{21}/2)$
and $\exp(i\alpha_{31}/2)$, respectively.}~$[0,4\pi]$ 
\cite{EMSPEJP09}
and the Dirac phase $\delta$ is varied 
in the interval $[0,2\pi]$. Shown 
are also the current experimental upper limit 
on the  $\mu\to e+\gamma$ decay 
branching ratio \cite{MEGA},
$B(\mu\to e+\gamma) < 1.2\times 10^{-11}$,
as well as the prospective limit 
of the MEG experiment \cite{MEG}, 
$B(\mu\to e+\gamma) < 10^{-13}$.

It is apparent from Fig.~\ref{fig1} that 
the data on the process
$\mu\to e + \gamma$ set very stringent 
constraints on the TeV scale see-saw mechanism. 
A neutrino Yukawa coupling $y=0.1$ 
generates a rate for the $\mu\to e+\gamma$ decay, 
which is ruled out by the MEGA experiment, unless
the RH neutrino mass is  $M_1\gtrsim 300$ GeV.
Furthermore, if the MEG experiment 
reaches the sensitivity of $10^{-13}$
without finding a signal, 
for the same Yukawa coupling $y=0.1$
the RH neutrino mass should 
be larger than 1 TeV.
More generally, 
for $M_1 = 100~{\rm GeV}$ ($M_1 = 1$ TeV) 
and $z \ll 1$ we get the following 
upper limit on the product 
$|(RV)_{\mu 1}^{*} (RV)_{e1}|$ 
of the heavy Majorana neutrino 
couplings to the muon (electron)
and the $W^\pm$ boson  
and to the $Z^0$ boson from the current 
upper limit \cite{MEGA}
on $B(\mu\to e+\gamma)$:
%%%%%%%%%%%%%%%%%%%%%%%%%%%%
\begin{equation}
\left |(RV)_{\mu 1}^{*}\, (RV)_{e1}\right| < 1.8\times 10^{-4}\,
(0.6\times 10^{-4})\,,
\label{T3}
\end{equation}
%%%%%%%%%%%%%%%%%%%%%%%%%%%
%
where we have used eqs.~(\ref{Bmutoeg1}) and (\ref{T3b}).
This can be recast as an upper bound on the 
neutrino Yukawa coupling $y$. Taking, $e.g.$ the best fit values of
the solar and atmospheric oscillation parameters, we get:
%%%%%%%%%%%%%%%%%%%%%%%%%%%%
\begin{eqnarray}
&& y\lesssim 0.036~(0.21)\,~{\rm for~NH~with~} M_1=100\,{\rm GeV}\,(1000\,{\rm GeV})~{\rm and~}\sin\theta_{13}=0 \,,\label{yupNH}\\
&& y\lesssim 0.031~(0.18)\,~{\rm for~IH~with~} M_1=100\,{\rm GeV}\,(1000\,{\rm GeV})~{\rm and~}\sin\theta_{13}=0\,,\label{yupIH}\\
&& y\lesssim 0.094~(0.54)\,~{\rm for~NH~with~} M_1=100\,{\rm GeV}\,(1000\,{\rm GeV})~{\rm and~}\sin\theta_{13}=0.2 \,,\label{yupNH0.2}\\
&& y\lesssim 0.16~(0.90)\,~{\rm for~IH~with~} M_1=100\,{\rm GeV}\,(1000\,{\rm GeV})~{\rm and~}\sin\theta_{13}=0.2\,.\label{yupIH0.2}
\end{eqnarray}
%%%%%%%%%%%%%%%%%%%%%%%%%%%%%
The upper limit on the neutrino Yukawa coupling  derived 
for $\theta_{13}=0$ is reached for 
$\alpha_{21}-\alpha_{31}\simeq \pi$ ($\alpha_{21}\simeq 3\pi$) in the case of
NH (IH) light neutrino mass spectrum. For $\sin\theta_{13}=0.2$,  
the upper limit on $y$
is obtained  for  $\alpha_{21}-\alpha_{31}\simeq \pi$ ($\alpha_{21}\simeq\pi$) 
and $\delta\simeq 0$ ($\delta\simeq0$),
if the neutrino masses have normal (inverted) hierarchy.

 A possible way to circumvent
these stringent upper bounds consists 
in assuming a very fine cancellation 
between the different terms in one of the 
eqs.~(\ref{Ue32NH})$-$(\ref{Umu21IH}) for the factors 
$|U_{\ell 3}+i\sqrt{m_{2}/m_{3}}U_{\ell 2}|^{2}$ 
and $|U_{\ell 2}+iU_{\ell 1}|^{2}$, $\ell=e,\mu$, 
in the expressions 
(\ref{meg-U-NH}) and (\ref{meg-U-IH})
for $B(\mu\to e+\gamma)$. 
Such a cancellation is possible in the case of NH spectrum (see also \cite{Strumia04}).
Indeed, we have $|U_{e3}+i\sqrt{m_{2}/m_{3}}U_{e 2}| = 0$ if   
%%%%%%%%%%%%%%%%%%%%%%%%%%%%%
\begin{equation}
\sin\left(\delta+\frac{\alpha_{21}-\alpha_{31}}{2}\right) = 1,~~
{\rm and}~~\tan\theta_{13}= 
\left(\frac{\dmsol}{\dma}\right)^{1/4}\,
\sin\theta_{12}\,. 
\label{Ue32NH0}
\end{equation}
%%%%%%%%%%%%%%%%%%%%%%%%%%%%%
%
In this case   $B(\mu\to e+\gamma) = 0$
and, thus the upper bound on the Yukawa 
coupling $y$ is no longer valid. 
Assuming the neutrino parameters 
have values within the corresponding present
$3\sigma$ ranges reported in Table~1, 
we find that the values of  $\theta_{13}$
implied by the condition in eq.~(\ref{Ue32NH0}) 
satisfy: $\tan\theta_{13}\gtrsim 0.21$, 
or $\sin^{2}\theta_{13}\gtrsim0.043$.
These values are outside 
the 3$\sigma$ interval of the experimentally 
allowed values of $\sin^{2}\theta_{13}$ (see Table 1).

We consider next the case in which the neutrino 
mass spectrum is with inverted hierarchy. 
The first thing to notice is that the factor
$| U_{e 2} + i U_{e 1}|^{2}$  can be rather small for 
$\sin (\alpha_{21}/2) = -1$ since
$2 c_{12}\,s_{12} \cong 0.93$, where we have used the 
best fit value of $\sin^2\theta_{12} = 0.312$.
In this case we have $| U_{e 2} + i U_{e 1}|^{2} \cong 0.069~(0.066)$   
for $\sin^2\theta_{13} = 0~(0.04)$.
Second, we show that, as in the case of normal 
hierarchical spectrum, it is possible to have a strong 
suppression of the factor $|U_{\mu 2}+iU_{\mu 1}|^{2}$  
if the CHOOZ mixing angle is close to the corresponding 
$3\sigma$ experimental upper bound. To be more 
quantitative, we take $\sin^{2}\theta_{12}=1/3$ and 
$\sin^{2}\theta_{23}=1/2$.
Then, expression (\ref{Umu21IH}) takes the form:
%%%%%%%%%%%%%%%%%%%%%%%%%%%%%%%%%%%%
\begin{eqnarray}
&&|U_{\mu 2}+iU_{\mu 1}|^{2}\;=\;\nonumber\\
&&\frac{1}{6}\left[2\left(\sqrt{2}(-1+s_{13}^{2}) - 
s_{13}\cos\delta\right)\sin\frac{\alpha_{21}}{2}+3\left(1+s_{13}^{2} + 
2s_{13}\cos\frac{\alpha_{21}}{2}\sin\delta\right)\right]\,.
	\label{Umu21IHb}
\end{eqnarray}
%%%%%%%%%%%%%%%%%%%%%%%%%%%%%%%%%%%%%
%
It is not difficult to show that, for fixed values of the phases 
$\alpha_{21}$ and $\delta$, $|U_{\mu 2}+iU_{\mu 1}|^{2}$ 
has a minimum for
%%%%%%%%%%%%%%%%%%%%%%%%%%%%%%%%%%%%
\begin{eqnarray}
\sin\theta_{13} &=& \frac{\cos\delta\sin\frac{\alpha_{21}}{2} - 
3\cos\frac{\alpha_{21}}{2}\sin\delta}
{3+2\sqrt{2}\sin\frac{\alpha_{21}}{2}}\,.
\label{s13min}
\end{eqnarray}
%%%%%%%%%%%%%%%%%%%%%%%%%%%%%%%%%%%%%
%
At the minimum, using eqs.~(\ref{Umu21IHb}) and (\ref{s13min}), 
we get:
%%%%%%%%%%%%%%%%%%%%%%%%%%%%%%%%%%%%
\begin{eqnarray}
{\rm min}\left(|U_{\mu 2}+iU_{\mu 1}|^{2}\right)&=&
\frac{\left(3\cos\delta\cos\frac{\alpha_{21}}{2} + 
\sin\delta\sin\frac{\alpha_{21}}
{2}\right)^{2}}{6\left(3 + 
2\sqrt{2}\sin\frac{\alpha_{21}}{2}\right)}\,.
\label{Umu21IHmin}
\end{eqnarray}
%%%%%%%%%%%%%%%%%%%%%%%%%%%%%%%%%%%
%
We will find next for which values of the 
CP violating phases $\delta$ and $\alpha_{21}$ 
this lower bound is equal to zero and 
if the resulting $\theta_{13}$, obtained from 
eq.~(\ref{s13min}), is compatible with
the existing limits from the neutrino oscillation data.
We have
${\rm min}(|U_{\mu 2}+iU_{\mu 1}|^{2})=0$ if
the Dirac and Majorana phases 
$\delta$ and $\alpha_{21}$ satisfy the 
following conditions: 
$\tan\delta \tan\frac{\alpha_{21}}{2}=-3$
and ${\rm sgn}(\cos\delta\cos\frac{\alpha_{21}}{2})= 
-{\rm sgn}(\sin\delta\sin\frac{\alpha_{21}}{2})$. 
Taking $\cos\delta>0$ ($\cos\delta<0$)
and using 
$\tan\delta= -3/ \tan(\alpha_{21}/2)$ 
in  eq. (\ref{s13min}) we get:
%%%%%%%%%%%%%%%%%%%%%%%%%%%%%%%%%%
\begin{equation}
\sin\theta_{13}\;=\, {\rm sgn}(\cos\delta) \, \frac{\sqrt{9 + 
\tan^{2}\frac{\alpha_{21}}{2}}}{3+2\sqrt{2}\sin\frac{\alpha_{21}}{2}}\,
\;\cos\frac{\alpha_{21}}{2}\,.
\label{s13min2}
\end{equation}
%%%%%%%%%%%%%%%%%%%%%%%%%%%%%%%%
%
The solution (\ref{s13min2})
is compatible with the $3\sigma$ upper limit of the 
CHOOZ mixing angle (see Table~1).
 In general,
one can always find a viable pair 
of CP violating phases $\alpha_{21}$ and $\delta$ 
satisfying the relations given above in order to set 
the r.h.s. of 
eq.~(\ref{Umu21IHmin}) equal to zero, if 
the mixing angle $\theta_{13}$ is sufficiently large, 
namely, if $\sin\theta_{13}> 3 - 2\sqrt{2} \cong 0.17$. 
More precisely, one finds, $e.g.$ that 
$|U_{\mu 2}+iU_{\mu 1}|^{2}\simeq 3.52\times10^{-8}~(2.43\times 10^{-6})$ 
for  $ s_{13}\simeq 0.2~(0.17)$, 
$\alpha_{21}\simeq 2.732~(\pi)$ and 
$\delta\simeq 5.725~(10^{-3})$.

%%%%%%%%%%%%%%%%%%%%%%%%%%%%%%%
%%%%%%%%%%%%%%%%%%%%%%%%%%%%%%%%%%%%
%

In order to interpret the results presented in 
Fig.~\ref{fig1}, it proves convenient to use 
the analytic expressions of 
$B(\mu\to e+\gamma)$ in terms of the low energy neutrino
parameters, the neutrino Yukawa coupling and the 
RH neutrino mass, eqs. (\ref{meg-U-NH})$-$(\ref{Umu21IH}).
Taking for concreteness $\sin^2 \theta_{23}\cong 1/2$,
$\sin^2 \theta_{12}\cong 1/3$ and using $\sin\theta_{13},
\Delta m^2_{\rm sol}/\Delta m^2_{\rm atm}\ll 1$,
eqs.~(\ref{meg-U-NH}) and (\ref{meg-U-IH}) approximately read:
%%%%%%%%%%%%%%%%%%%%%%%%%%%%%%%%%%%%%
\begin{eqnarray}
&&{\rm\bf NH:}\;\;\;B(\mu\to e+\gamma)\cong \frac{3\alpha_{\rm em}}{32\pi}\left(\frac{y^{2}v^{2}}{M_{1}^{2}}\right)^{2}\,\left[G(X)-G(0) \right]^{2}
 \frac{1}{6}\left(\frac{\dmsol}{\dma}\right)^{1/4}\times\nonumber \\\\
&& 
\left|\left[\left(\frac{\dmsol}{\dma}\right)^{1/4}
-2\sqrt{3}\sin\theta_{13}
\sin\left(\delta+\frac{\alpha_{21}-\alpha_{31}}{2}
\right)\right]\left[1-2\sqrt{\frac{2}{3}}\left(\frac{\dmsol}{\dma}\right)^{1/4}
\sin\left(\frac{\alpha_{21}-\alpha_{31}}{2}\right)\right]
\right|\,,\nonumber\\\nonumber&&\\
&&{\rm\bf IH:} \;\;\;B(\mu\to e+\gamma)\cong \frac{3\alpha_{\rm em}}{32\pi}
\left(\frac{y^{2}v^{2}}{M_{1}^{2}}\frac{1}{2}\right)^{2}\,\left[G(X)-G(0) \right]^{2}\times\nonumber\\\\
&&\frac{1}{18}\left|5+4\cos\alpha_{21}-2\sin\theta_{13}\left(3+2\sqrt{2}\sin\frac{\alpha_{21}}{2}\right)
\left(\cos\delta\sin\frac{\alpha_{21}}{2}-3\sin\delta\cos\frac{\alpha_{21}}{2}\right)\right|
\,.\nonumber
\end{eqnarray}
%%%%%%%%%%%%%%%%%%%%%%%%%%%%%%%%%%%
%
From these expressions it follows that there is a fairly strong 
dependence of the prediction of $B(\mu\to e+\gamma)$ on the
Majorana phases. As a result, and since there is no proposal
to constrain experimentally these phases, there is uncertainty band of a 
factor of five for normal hierarchy and a factor of nine
for inverted hierarchy, which cannot be reduced even if 
the neutrino masses and mixing angles were known with
arbitrarily high precision.

  It is not difficult to get also expressions for the 
double ratios 
$R(21/31) =B(\mu\to e+\gamma)/B'(\tau\to e +\gamma)$ and 
$R(21/32) =B(\mu\to e+\gamma)/B'(\tau\to \mu +\gamma)$,
where $B'(\tau\to e(\mu) +\gamma) \equiv 
B(\tau\to e(\mu) +\gamma)/B(\tau\to e(\mu)+ \nu_{\tau}+\bar{\nu}_{e(\mu)})$,
$B(\tau\to e(\mu) +\gamma)$ and 
$B(\tau\to e(\mu)+ \nu_{\tau}+\bar{\nu}_{e(\mu)})$ being 
the branching ratios of the corresponding decays.
The double ratios of interest depend only 
on the neutrino masses and the elements of the PMNS matrix:
%%%%%%%%%%%%%%%%%%%%%%%%%%%%
\begin{eqnarray}
&&R(21/31)
\cong
\frac{\left|U_{\mu 3}+i\sqrt{\frac{m_{2}}{m_{3}}}U_{\mu 2} \right|^{2}}
{\left|U_{\tau 3}+i\sqrt{\frac{m_{2}}{m_{3}}}U_{\tau 2} \right|^{2}}\,,
~{\rm NH}\,,\\
&&R(21/31)\cong \frac{\left|U_{\mu 2}+iU_{\mu 1} \right|^{2}}
{\left|U_{\tau 2}+iU_{\tau 1} \right|^{2}}\,,~{\rm IH}\,;
\end{eqnarray}
%%%%%%%%%%%%%%%%%%%%%%%%%%%
%
%
%%%%%%%%%%%%%%%%%%%%%%%%%%%%
\begin{eqnarray}
&&R(21/32)
\cong
\frac{\left|U_{e 3}+i\sqrt{\frac{m_{2}}{m_{3}}}U_{e 2} \right|^{2}}
{\left|U_{\tau 3}+i\sqrt{\frac{m_{2}}{m_{3}}}U_{\tau 2} \right|^{2}}\,
~{\rm NH}\,,\\
&&R(21/32)\cong
\frac{\left|U_{e 2}+iU_{e 1} \right|^{2}}
{\left|U_{\tau 2}+iU_{\tau 1} \right|^{2}}\,,~{\rm IH}\,.
\end{eqnarray}
%%%%%%%%%%%%%%%%%%%%%%%%%%%
% 
For the NH (IH) light neutrino mass spectrum,
the range of variability at $3\sigma$ of 
each of the two ratios defined above is:
$0.01 \lesssim R(21/31)\lesssim 20$ 
($0.001 \lesssim R(21/31)\lesssim 300$) and 
$5\times10^{-4} \lesssim R(21/32)\lesssim 3$ 
($0.04 \lesssim R(21/32)\lesssim 5000$).~\footnote{
In order to get such estimates we assume that the neutrino observables have  
a gaussian distribution with the corresponding mean value and standard deviation reported
in Table 1.}~Thus, in the case of NH spectrum, the predicted
$\tau\to \mu +\gamma$ decay branching ratio
$ B'(\tau\to \mu +\gamma)$ in the scheme considered 
can be by several orders of magnitude larger than the
$\mu\to e+\gamma$ decay branching ratio,
$B(\mu\to e+\gamma)$. 

\begin{table}[t]\centering
    \begin{tabular}{|c|cc|cc|}
        \hline
        Parameter & \multicolumn{2}{c|}{Best Fit $\pm\,1\sigma$}& \multicolumn{2}{c|}{3$\sigma$ interval} 
        \\ \hline\hline
        &\multicolumn{1}{c|}{NH}& IH& \multicolumn{1}{c|}{NH} & IH\\ \cline{2-5}
        &\multicolumn{2}{c|}{}&\multicolumn{2}{c|}{}\\
        $\Delta m^2_{21}\: (10^{-5}{\rm eV}^{2})$
        & \multicolumn{2}{c|}{$7.59^{+0.20}_{-0.18}$}    &\multicolumn{2}{c|}{$[7.09,8.19]$} \\[3mm]
        $|\Delta m^2_{31}|\: (10^{-3}{\rm eV}^{2})$
        & $2.45^{+0.09}_{-0.09}$  & $2.34^{+0.10}_{-0.09}$ & $[2.18,2.73]$ & $[2.08,2.64]$ \\[3mm]
        $\sin^2\theta_{12}$
        & \multicolumn{2}{c|}{$0.312^{+0.017}_{-0.015}$} &\multicolumn{2}{c|}{$[0.27,0.36]$} \\[3mm]  
        $\sin^2\theta_{23}$
        & $0.51\pm 0.06$ &$0.52\pm 0.06$ & \multicolumn{2}{c|}{$[0.39,0.64]$}\\[3mm] 
        $\sin^2\theta_{13}$
        & $0.010^{+0.009}_{-0.006}$  & $0.013^{+0.009}_{-0.007}$& $\leq$ 0.035 & $\leq$ 0.039 \\[3mm]
        \hline
\end{tabular}
\begin{center}
\caption{ \label{tab:mixing2010} Best fit values with 1$\sigma$ errors and
3$\sigma$ intervals for the three flavour neutrino  oscillation parameters (see \cite{Schwetz:2008er} 
and references therein).} 
\end{center}
\end{table}

%%%%%%%%%%%%%%%%%%%%%%%%%%%%%%%
\begin{figure}
\begin{center}
\begin{tabular}{cc}
\includegraphics[width=7.5cm,height=6.5cm]{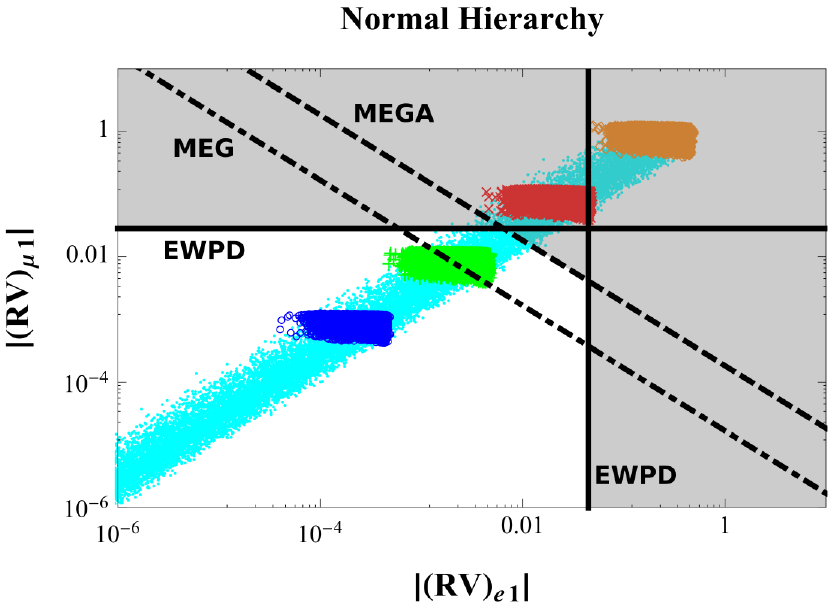} &
\includegraphics[width=7.5cm,height=6.5cm]{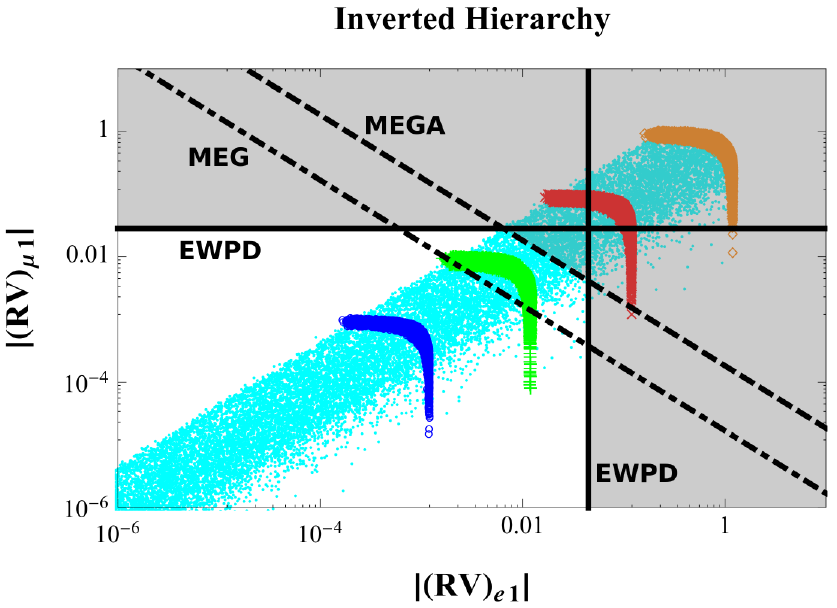}\\
\includegraphics[width=7.5cm,height=6.5cm]{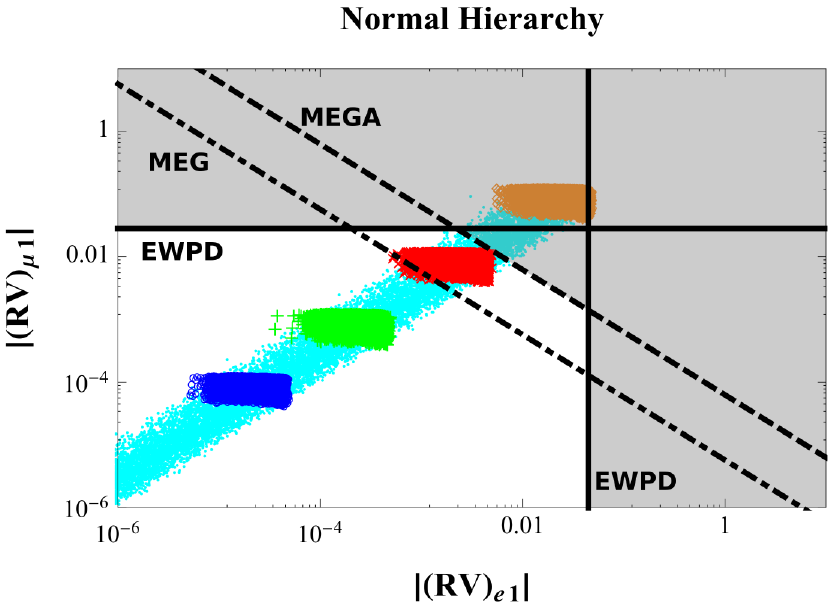} &
\includegraphics[width=7.5cm,height=6.5cm]{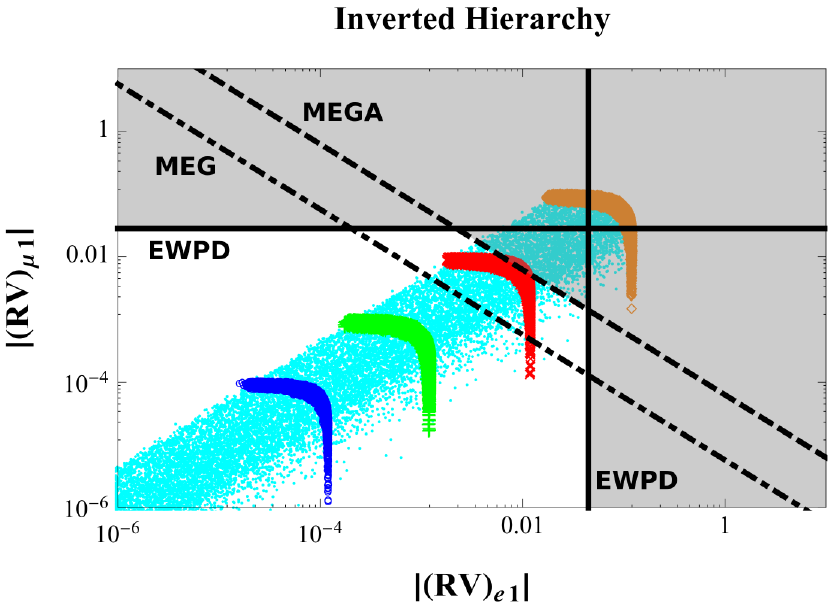}
\end{tabular}
\caption{Correlation between $|(RV)_{e1}|$ and $|(RV)_{\mu1}|$ in the case of NH (left panels)
and IH (right panels) light neutrino mass spectrum, for 
{ $M_{1}=100\,(1000)$ GeV, upper (lower) panels, $i)$ $y=0.001$ (blue $\circ$), $ii)$
$y=0.01$ (green $+$), $iii)$ $y=0.1$ (red $\times$) and $iv)$ $y=1$ (orange $\Diamond$)}. 
The cyan points correspond
to random values of $y\leq 1$. The dashed line corresponds to the MEGA
bound \cite{MEGA}, $B(\mu\rightarrow e +\gamma)\leq
1.2 \times 10^{-11}$. The dot-dashed line
corresponds to $B(\mu\rightarrow e +\gamma) =
10^{-13}$, which is the prospective sensitivity
of the MEG experiment \cite{MEG}.\label{fig2}}
\end{center}
\end{figure}
%%%%%%%%%%%%%%%%%%%%%%%%%%%%%%%%%%%%

\section{Implications for Collider Searches and Electroweak 
Precision Observables}

%%%%%%%%%%%%%%%%%%%%%%%%%%%%%%%%

  Upper bounds on the couplings of RH neutrinos with SM particles can be 
set by analyzing lepton number conserving processes like
$\pi\to\ell \overline{\nu}_{\ell}$, $Z\to \nu\overline{\nu}$ 
and other tree-level processes involving light neutrinos in the final
state \cite{Antusch:2008tz}. 
From eqs.~(\ref{eta}) and (\ref{eta_bounds}), we get:
\begin{eqnarray}
 |(RV)_{e1}|^{2} & \lesssim & 2\times 10^{-3}
\label{e-bound}\,,\\
 |(RV)_{\mu 1}|^{2} &\lesssim & 0.8\times 10^{-3}
\label{mu-bound}\,,\\
|(RV)_{\tau 1}|^{2} & \lesssim & 2.6\times 10^{-3}
\label{tau-bound}\,.	
\end{eqnarray} 
%%%%%%%%%%%%%%%%%%%%%%%%%%%%%%%
%
Following the same rationale as in the previous section, one can
translate the upper bounds on the RH neutrino couplings
from electroweak precision observables
into upper bounds on the neutrino Yukawa coupling $y$. 
 For this purpose we observe that
an upper limit on $y$, which is independent on the specific values of the
neutrino oscillation parameters, can be easily derived from eq.~(\ref{ymax2}), taking into account the 
experimental constraints given above:
\begin{equation}\label{yewpd}
	y\;\lesssim\; 0.06\,\left(\frac{M_{1}}{100~{\rm GeV}}\right)\,.
\end{equation}
 
 On the other hand, using
eq.~(\ref{mixing-vs-yIH})  and taking the best fit 
values of neutrino oscillation parameters, we obtain 
%%%%%%%%%%%%%%%%%%%%%%%%%%%%%%%%%%%%%%%%%%%%
\begin{eqnarray}
&& y\lesssim 0.047~(0.47)\,~{\rm for~NH~with~} M_1=100\,{\rm GeV}\,(1000\,{\rm GeV})~{\rm and~}\sin\theta_{13}=0 \,,\label{yupNHb}\\ 
&& y\lesssim 0.046~(0.46)\,~{\rm for~IH~with~} M_1=100\,{\rm GeV}\,(1000\,{\rm GeV})~{\rm and~}\sin\theta_{13}=0\,,\label{yupIHb}\\ 
&& y\lesssim 0.049~(0.49)\,~{\rm for~NH~with~} M_1=100\,{\rm GeV}\,(1000\,{\rm GeV})~{\rm and~}\sin\theta_{13}=0.2 \,,\label{yupNH0.2b}\\
&& y\lesssim 0.053~(0.53)\,~{\rm for~IH~with~} M_1=100\,{\rm GeV}\,(1000\,{\rm GeV})~{\rm and~}\sin\theta_{13}=0.2\,.\label{yupIH0.2b}
\end{eqnarray}
%%%%%%%%%%%%%%%%%%%%%%%%%%%%%%%%%%%%%%%%%%%%
%
 The results reported in eqs.~(\ref{yupNHb}) and (\ref{yupIHb}) 
are obtained for $\alpha_{21}-\alpha_{31}\simeq \pi$
and $\alpha_{21}\simeq0$, respectively. 
In the case of $\sin\theta_{13}=0.2$ 
and NH (IH) light neutrino mass spectrum, the upper bound on $y$ 
corresponds to $\alpha_{21}-\alpha_{31}\simeq \pi$ 
($\alpha_{21}\simeq 0$) and $\delta\simeq \pi$ ($\delta\simeq 3\pi/2$).

  The bounds for $\sin\theta_{13}=0$ are {\it weaker} 
than those derived in the previous
section from the non-observation of the 
$\mu\rightarrow e+\gamma$ decay.
Thus, the existing stringent upper bound 
on the  $\mu\rightarrow e +\gamma$ decay
rate makes very difficult the observation 
of deviations from the Standard
Model in the electroweak precision data, predicted 
by the TeV scale see-saw scenario.
In contrast, in the case of 
$\sin\theta_{13}=0.2$, the constraint from 
the MEGA upper bound \cite{MEGA} can be avoided  
and we get a better upper limit on $y$  
from the electroweak precision data. 
More precisely, we find that 
for $M_{1}=100~(1000)$ GeV, the limits given 
in eqs. (\ref{e-bound})$-$(\ref{tau-bound}) provide a better
constraint on the neutrino Yukawa coupling $y$
than the upper bound on $B(\mu \to e + \gamma)$ 
if $\sin\theta_{13}>0.10~(0.19)$ in the case of 
the NH neutrino mass spectrum,
and for $\sin\theta_{13}>0.13~(0.17)$ if the 
spectrum is of the IH type.
 In particular,
the parameter $y$ can have a value
as large as  $4\pi$ 
that is still compatible with 
the current upper limit on
$B(\mu \to e + \gamma)$ \cite{MEGA}. 
This is possible in the case of 
NH (IH) light neutrino mass spectrum
for sufficiently large values of 
$\sin\theta_{13}\gtrsim 0.22~(0.18)$. 
We note, however, that 
at $\sin\theta_{13} = 0.10~(0.19)$ 
($\sin\theta_{13} = 0.13~(0.17)$)
for the NH (IH) spectrum and 
$M_{1}\approx 100~(1000)$ GeV, 
we have $y\ltap 0.05~(0.5)$  
both from the bound on $B(\mu \to e + \gamma)$ 
and the electroweak  data limits. This is consistent with
the absolute upper limit reported in (\ref{yewpd}).

  These results are illustrated in Fig.~\ref{fig2}, 
where we show, for $M_1=100\,{\rm GeV}$  (upper panels) and 
$M_{1}=1000$ GeV (lower panels), the
allowed ranges of the RH
neutrino couplings $|(RV)_{\mu1}|$ and $|(RV)_{e 1}|$, 
in the case of NH (left panels) and IH (right panels) spectra. 
The region of the parameter space 
which is allowed by the electroweak precision data, 
eqs.~(\ref{e-bound}) and (\ref{mu-bound}),
is indicated with solid lines; the region allowed
by the current bound on the $\mu\rightarrow e+\gamma$ 
decay rate is indicated with a dashed line.
The projected MEG sensitivity reach is shown with
a dot-dashed line. 
Finally, in the same figure we show  
a scatter plot of the points 
which are consistent with the $3\sigma$ allowed
ranges of the neutrino oscillation parameters.
This is done for four different values of 
the neutrino Yukawa coupling $y$: 
$y=0.001$ (blue $\circ$), $ii)$
$y=0.01$ (green $+$), $iii)$ $y=0.1$ (red $\times$) and $iv)$ $y=1$ (orange $\Diamond$). 
The cyan points in each plot correspond to
an arbitrary value of the Yukawa coupling $y \leq 1$.

  After imposing all the constraints on the parameter
space discussed above, we find a fairly narrow 
band of values of  $|(RV)_{\mu1}|$ and $|(RV)_{e 1}|$, 
which is allowed by the data, more precisely,
by the requirement of reproducing the correct 
values of the neutrino oscillation parameters 
and by the constraint following from the 
upper bound on the $\mu\rightarrow e+\gamma$ decay 
rate. Interestingly, 
and as we have already mentioned earlier, 
the limits from the electroweak precision data 
lie essentially in the excluded (shaded) region. 
Therefore, in the view of the
constraints from the data on the neutrino 
oscillation parameters and the $\mu \to e + \gamma$ decay,
the discovery of significant deviations 
from the Standard Model predictions
in the electroweak precision observables 
appears highly improbable, 
unless a significant improvement 
in the precision of the data is achieved.
The above conclusion will be strengthen
if the MEG experiment reaches the
sensitivity $B(\mu\rightarrow e\gamma)\sim 10^{-13}$ 
without observing a signal.

It has been argued that present constraints 
from electroweak precision
data allow the observation of pseudo-Dirac RH neutrinos
at the LHC. More concretely, the following process with 
three charged lepton final state,
%%%%%%%%%%%%%%%%%%%%%%%%%%%%%%%%%%%%%
\begin{equation}
q\overline{q}^{\prime}\to \mu^{+} N_{PD}\to \mu^{+}\mu^{-} W^{+}\to\mu^{+}\mu^{-}\mu^{+}\nu_{\mu}\,.\label{3muons}
\end{equation}
%%%%%%%%%%%%%%%%%%%%%%%%%%%%%%%%%%%%
%
could be observed at the LHC with a luminosity of 13 ${\rm fb}^{-1}$ 
with a  $5\sigma$ significance if the RH neutrinos have a mass of
$\sim 100\,{\rm GeV}$~\cite{delAguila:2008hw}. 
A discovery reach in this channel implies a 
coupling $|(RV)_{\mu 1}|\approx0.04$  
\cite{delAguila:2008hw}.
Taking $|(RV)_{\mu1}|\gtrsim 0.04$, 
we get from (\ref{mixing-vs-y}) and (\ref{mixing-vs-yIH}):
%%%%%%%%%%%%%%%%%%%%%%%%%%%%%
\begin{eqnarray}
&& y\gtrsim 0.04\,~{\rm for~NH~with~} M_1=100\,{\rm GeV}\,, \\
&& y\gtrsim 0.05\,~{\rm for~IH~with~} M_1=100\,{\rm GeV}\,.	
\end{eqnarray}
%%%%%%%%%%%%%%%%%%%%%%%%%%%%%%%
%
which should be compared with
the upper limits on $y$ we get from 
the upper bound on
the $\mu\to e +\gamma$ decay rate, 
eqs.~(\ref{yupNH}) and (\ref{yupIH}), 
or from electroweak precision data, 
eqs.~(\ref{yupNH0.2b}) and (\ref{yupIH0.2b}). 
We find again a tension between the  
constraints on $y$ obtained from the 
data on the $\mu\rightarrow e+\gamma$ 
decay or the electroweak processes 
and the values of $y$ required for
the production of RH neutrinos
with observable rates at colliders.
This is clearly seen in Fig.~\ref{fig2}. 

 As a conclusion, the presently existing data on the 
neutrino mixing parameters and
the present experimental upper bound 
$B(\mu\to e+\gamma)<1.2\times 10^{-11}$,
basically rule out the possibility of producing pseudo-Dirac
neutrinos at LHC with observable rates. A similar conclusion
applies to the possibility of observing deviations to the Standard Model
predictions in the electroweak precision observables. 
An improvement of the bound on the $\mu\rightarrow e+\gamma$
decay rate by a factor of $\sim 10$ will make the 
observation of pseudo-Dirac 
neutrinos at the LHC completely impossible and
the effects on the electroweak precision observables negligibly small,
unless the RH neutrinos have additional (flavour-conserving)
couplings to the Standard Model particles, 
as like in theories with  extra $U(1)$ local gauge symmetry 
under which the pseudo-Dirac neutrinos are charged,
or the TeV scale type III see-saw scenario.

%%%%%%%%%%%%%%%%%%%%%%%%%%%%%%%%%%%%%%%
%
\mathversion{bold}
\section{Predictions for the \betabeta-Decay}
\mathversion{normal}

In the scenario we are considering, 
the $\betabeta$-decay effective Majorana 
mass $\meff$,
which controls the \betabeta-decay rate,
receives a contribution from the exchange of 
the heavy Majorana neutrino fields $N_{k}$ \cite{HPR83}, 
which may be not negligible for ``large'' neutrino 
Yukawa couplings. One has, in general:
%%%%%%%%%%%%%%%%%%%%%%%%%%%%%%%%%%%%%%
\begin{equation} 
\meff \cong 
\left |\sum_{i=1}^{3}U^2_{ei}\, m_i 
- \sum_{k=1}^{2}\, F(A,M_k)\, (RV)^2_{e k}\,M_k \right |\,,
\label{mee1}
\end{equation}
%%%%%%%%%%%%%%%%%%%%%%%%%%%%%%%%%
%
where~\footnote{Let us note that the interference between the contributions
due to the light and heavy Majorana neutrino exchanges
is not suppressed because both contributions
are generated by the weak interaction involving
currents of the same $(V - A)$ structure.
The interference term of interest would be
strongly suppressed if the heavy Majorana neutrino
exchange is generated by $(V + A)$
currents \cite{HPR83}.}~ the function $F(A,M_{k})$ depends on the 
$N_k$ masses and the 
type of decaying nucleus $(A,Z)$. For
$M_{k}=(100\div1000)$ GeV, one can use 
the rather accurate approximate expression 
for $F(A,M_k)$ \cite{JV83}: 
$F(A,M_k)\cong(M_{a}/M_{k})^{2}f(A)$, 
where $M_{a}\approx 0.9$ GeV and  $f(A)$
depends on the decaying isotope considered. 
For, e.g.,  $^{76}$Ge, $^{82}$Se, $^{130}$Te and $^{136}$Xe, 
the function $f(A)$ takes the values $f(A)\cong$ 
0.079, 0.073, 0.085 and 0.068, respectively. 
In the case of $^{48}$Ca, $f(A)$ has a smaller value 
\cite{JV83}: $f({\rm ^{48}Ca}) \cong 0.033$. 
Using eq.~(\ref{rel1}) we can write the 
heavy Majorana neutrino exchange contribution 
to $\meff$ in a simplified form:
%%%%%%%%%%%%%%%%%%%%%%%%%%%%%%%%%%
\begin{equation}
\mefff^{{\rm N}} 
\cong - \,\frac{2z + z^2}{(1 + z)^2}\,
\left(RV\right)_{e1}^{2}\, \frac{M_{a}^{2}}{M_{1}}\,f(A)\,.
\label{mee3}
\end{equation}
%%%%%%%%%%%%%%%%%%%%%%%%%%%%%%%%%%
%
This contribution, as we will show below,
can be even as large as $|\mefff^N|\sim 0.2~(0.3)$ eV 
in the case of NH (IH) light neutrino 
mass spectrum.~\footnote{Note that in the approximation we use 
for $(RV)_{e k}$ one has $\sum_{k=1}^{2}\,(RV)^2_{e k}\,M_k = 0$.
The heavy Majorana neutrino exchange 
contribution $\mefff^N$ to $\meff$
is not zero due to the nontrivial dependence on $M_k$ of 
the function  $F(A,M_k)$, see eq. (\ref{mee1}).}

\begin{figure}
\begin{center}
\begin{tabular}{cc}
\includegraphics[width=7cm,height=6.5cm]{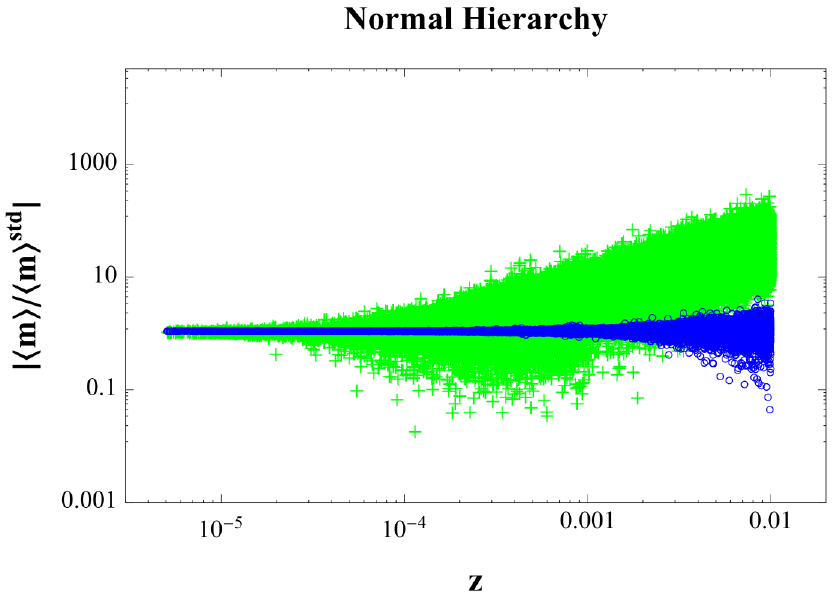} &
\includegraphics[width=7cm,height=6.5cm]{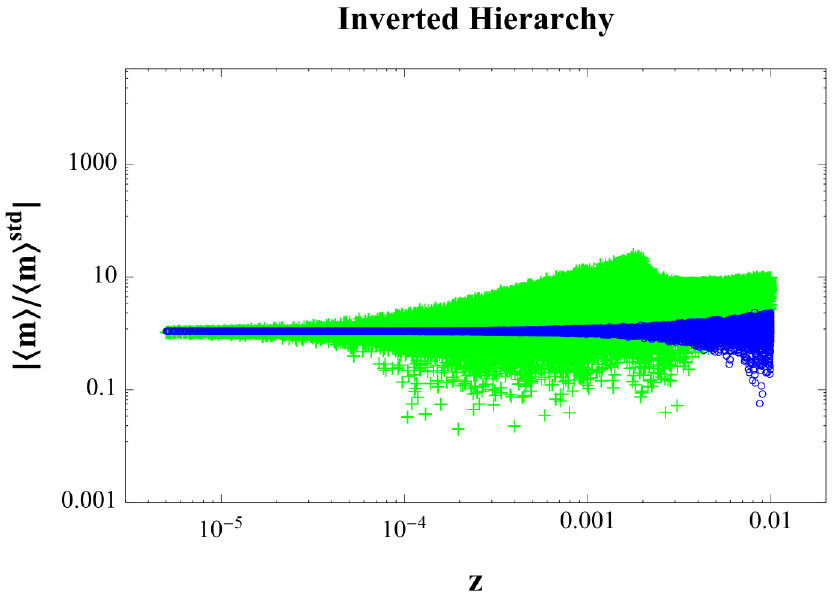}
\end{tabular}
\caption{The ratio between the effective Majorana mass 
$\meff$ and the ``standard'' contribution 
$|\mefff^{{\rm std}}|$ for $^{76}$Ge
in the cases of NH (left panel)
and IH (right panel) light neutrino mass spectrum, 
for $M_{1}=100$ GeV and $i)$ $y=0.001$ (blue $\circ$), $ii)$
$y=0.01$ (green $+$). 
\label{fig3}}
\end{center}
\end{figure}

\begin{figure}
\begin{center}
\begin{tabular}{cc}
\includegraphics[width=7.5cm,height=6.5cm]{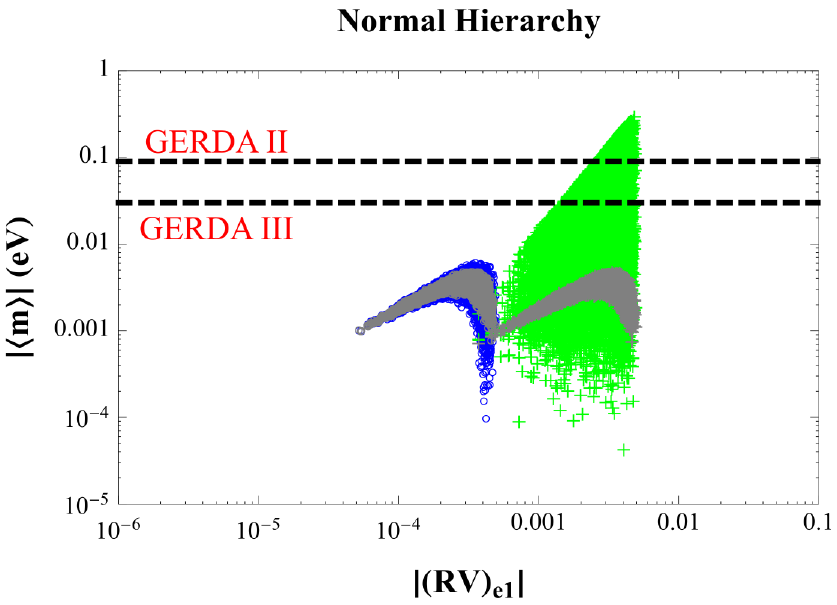} &
\includegraphics[width=7.5cm,height=6.5cm]{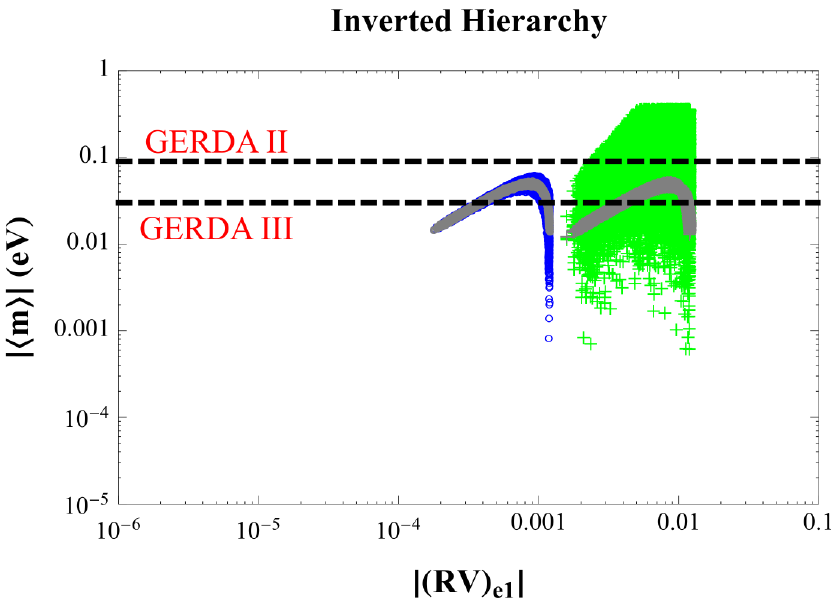}
\end{tabular}
\caption{The effective Majorana mass 
$\meff$ vs $|(RV)_{e1}|$ for $^{76}$Ge in the cases of NH (left panel)
and IH (right panel) light neutrino mass spectrum,
for $M_{1}=100$ GeV and $i)$ $y=0.001$ (blue $\circ$), $ii)$
$y=0.01$ (green $+$). 
The gray markers correspond to $|\mefff^{{\rm std}}|$.
\label{fig4}}
\end{center}
\end{figure}

   In Fig.~\ref{fig3} we show the ratio between 
the total effective Majorana mass $\meff$
given by eq.~(\ref{mee1}) and the ``standard'' 
contribution due to the light Majorana neutrino exchange 
(see, e.g., \cite{BPP1,bb0nuNHIH}):  
$|\mefff^{{\rm std}}|\equiv |\sum_{i=1}^{3}U^2_{ei}\, m_i |$. 
In this plot we considered a nuclear matrix element factor $f(A)$
corresponding to $^{76}$Ge, although the conclusions are analogous
for other nuclei. The scan of the parameter space was done in 
the same way as in Fig.~\ref{fig1}, selecting just the points 
which are in agreement with the present experimental bound on 
$B(\mu\rightarrow e+\gamma)$.
In Fig.~\ref{fig4} we show the range of the 
possible values of $\meff$ as function of $|(RV)_{e1}|$
for $M_{1}=100$ GeV and $z<10^{-2}$,
in the case of \betabeta-decay of $^{76}$Ge.

  The results of the analysis illustrated graphically  
in Figs.~\ref{fig3} and \ref{fig4} demonstrate
that RH neutrinos can significantly
enhance the rate of $\betabeta$-decay, 
without this being in conflict with 
the present upper bound on the 
$\mu\rightarrow e+\gamma$ decay rate. 
For example,
$\meff$ can be as large as  $0.2~(0.3)$ eV,
if $z\cong 10^{-3}\,(10^{-2})$
and $M_{1}\approx 100\, (1000)$ GeV.
In contrast, in the limit $z\rightarrow 0$,
the RH neutrinos behave as a Dirac pair and hence do not
contribute to the $\betabeta$-decay amplitude
to leading order.
 
  Let us recall that in the scheme with 
two heavy Majorana neutrinos we are considering,
the lightest neutrino mass is zero. This implies 
that for the NH and IH light neutrino mass spectra 
we have for the ``standard'' contribution 
to $\meff$, respectively 
(see, e.g., \cite{bb0nuNHIH}):
$|\mefff^{{\rm std}}| \ltap 0.005$ eV and
$0.01~{\rm eV} \ltap |\mefff^{{\rm std}}| \ltap 0.05$ eV.
Thus, if  it is established that the 
light neutrino mass spectrum is hierarchical, 
a value of $\meff > 0.05$ eV would signal 
the presence of a contribution to 
$\meff$ beyond the ``standard'' one.
In the scheme considered,   $\meff$ 
can be by a factor up to $\sim 100$ ($\sim 10$) 
larger than the maximal value of $|\mefff^{{\rm std}}|$ 
predicted in the case of NH (IH)
light neutrino mass spectrum.

  It should be noted also that 
the predicted value of 
$\meff$ in the cases of the 
NH (IH) spectrum exhibits a 
strong dependence on the PMNS 
parameters and especially
on the Majorana phase \cite{BHP80}
$\alpha_{21} - \alpha_{31}$ 
($\alpha_{21}$). If
we have  $|\mefff^{{\rm std}}|\cong |\mefff^{{\rm N}}|$, 
i.e.,  if the ``standard'' contribution to 
$\meff$ is of the same order 
as the contribution  due to the exchange of 
the heavy Majorana neutrinos $N_{1,2}$ 
(see eq.~(\ref{mee3})), 
$\meff$ will depend also on the phase $\omega$
(see eq. (\ref{RVpseudo})). 

Finally, if for a given decaying 
nucleus $(A_1,Z_1)$ and certain values 
of the parameters of the problem 
there is a strong mutual 
compensation between the two 
contributions $\mefff^{{\rm std}}$ and $\mefff^{{\rm N}}$
in $\meff$ and we have 
$\meff \ll |\mefff^{{\rm std}}|,|\mefff^{{\rm N}}|$,
similar cancellation will not happen, in general, 
for another decaying nucleus $(A_2,Z_2)$ 
due to the dependence of 
$\mefff^{{\rm N}}$ on $(A,Z)$ \cite{HPR83}.
In the scheme considered, in which only
hierarchical light neutrino mass 
spectrum is possible, 
and in view of the planned 
sensitivity of the next generation of 
$\betabeta$-decay experiments,
this observation has practical importance
if the light neutrino mass spectrum is 
with inverted hierarchy.~In a more general context 
in which the 3rd heavy Majorana neutrino 
is relevant for the see-saw mechanism and 
the light neutrino mass spectrum can be 
with partial hierarchy or even of the 
quasi-degenerate type (see, e.g., \cite{BPP1,PDG10}),
the observation made above regards both 
types of light neutrino mass spectrum - 
with normal ordering and with inverted ordering.
Indeed, $\mefff^{{\rm N}}$ in the general case 
can be written in the form:
%%%%%%%%%%%%%%%%%%%%%%%%%%%%%%%%%%%%%%
\begin{equation}
 \mefff^{{\rm N,A}} \cong - f(A)\,\tilde{F}^{N}\,,~~
\tilde{F}^{N}\equiv \sum_{k=1}^{3}\, (RV)^2_{ek}\,\frac{M^2_{a}}{M_{k}}\,,
\label{meeN2}
\end{equation}
%%%%%%%%%%%%%%%%%%%%%%%%%%%%%%%%%
%
where $\tilde{F}^{N}$ does not depend on $A$.
Suppose that for a given isotope $(A_1,Z_1)$ the 
two terms in $\meff$,  $\mefff^{{\rm std}}$ and
$\mefff^{{\rm N,A_1}}$, mutually compensate 
each other so that 
$|\mefff^{A_1}| \ll |\mefff^{{\rm std}}|,|\mefff^{{\rm N,A_1}}|$.
That would imply that 
%%%%%%%%%%%%%%%%%%%%%%%%%%%%%%%%%%%%%%
\begin{equation}
 \mefff^{{\rm N,A_1}} \equiv f(A_1)\,\tilde{F}^{N}
\cong  \sum_{i=1}^{3}U^2_{ei}\, m_i\,. 
\label{meeNA1}
\end{equation}
%%%%%%%%%%%%%%%%%%%%%%%%%%%%%%%%%
%
In this case the effective Majorana mass 
corresponding to an isotope 
$(A_2,Z_2)$ will be given by:
%%%%%%%%%%%%%%%%%%%%%%%%%%%%%%%%%%%%%%
\begin{equation}
 |\mefff^{A_2}| \cong \left |1 - \frac{f(A_2)}{f(A_1)}\right |\,
\left | \sum_{i=1}^{3}U^2_{ei}\, m_i\right | \,. 
\label{meeNA2}
\end{equation}
%%%%%%%%%%%%%%%%
% 
If, e.g., the cancellation between the two terms 
in $\mefff^{A_1}$ occurs for  $^{48}$Ca (for which 
 $f({\rm ^{48}Ca}) \cong 0.033$), it will not 
take place for  e.g.,  $^{76}$Ge, $^{82}$Se, $^{130}$Te and $^{136}$Xe
since for these isotopes  $f(A_2)\cong$ 
0.079, 0.073, 0.085 and 0.068, respectively.
Actually, for $^{76}$Ge, $^{82}$Se and $^{130}$Te
the factor $|1 - f(A_2)/f(A_1)|\cong$ 1.39, 1.21 and 1.58,
so we will have a somewhat larger  $ |\mefff^{A_2}|$
than the ``standard'' one $|\mefff^{{\rm std}}|$, 
while for  $^{136}$Xe the indicated factor 
is  1.06 and thus $ |\mefff^{A_2}|\cong 
|\mefff^{{\rm std}}|$. If, however, the 
cancellation between the two terms in 
 $\mefff^{A_1}$ takes place for, e.g., 
one of the nuclei  $^{76}$Ge, $^{82}$Se, 
$^{130}$Te and $^{136}$Xe, for which 
the function $f(A)$ has rather similar 
values, $|\mefff^{A_2}|$ for the other nuclei
will be suppressed with respect 
to $|\mefff^{{\rm std}}|$ to various degrees.
For instance, if the cancellation is 
operative for $^{136}$Xe, 
$|\mefff^{A_2}|$ for  $^{76}$Ge, $^{82}$Se, $^{130}$Te and 
$^{48}$Ca will be suppressed with respect to the 
``standard'' contribution 
 $|\mefff^{{\rm std}}|$ by the factors
0.16, 0.07, 0.25 and 0.51, respectively.

In the case with two heavy Majorana neutrinos 
we are considering and for IH light neutrino mass spectrum,
the condition for an exact cancellation between 
$\mefff^{{\rm std}}$ and $\mefff^{{\rm N,A}}$ can be 
easily derived in terms of the basic parameters of the scheme.
Using eqs.~(\ref{mixing-vs-y}), (\ref{mixing-vs-yIH}),  (\ref{mee1}) and (\ref{mee3}) we
can write $\meff$ as:
%%%%%%%%%%%%%%%%%%%%%%%%%%%%%%%%
\begin{eqnarray}
\meff &\simeq& \left| m_{1} U_{e1}^{2}\left( 1 -K \right)\,+
\,m_{2} U_{e2}^{2}\left( 1 + K \right)\,+
\,2i\sqrt{m_{1} m_{2}}K  \left(U_{e2}U_{e1}\right)\right|\,,
\end{eqnarray} 
%%%%%%%%%%%%%%%%%%%%%%%%%%
%
where $K$ is given by:
%%%%%%%%%%%%%%%%%%%%%%%%%%%%%%%%%%%
\begin{equation}
K\;\simeq\; \frac{z}{2} \frac{y^{2} v^{2}}{M_{1}^{2}}\frac{M_{a}^{2}}
{M_{1}\sqrt{\dma}}f(A)e^{-2i\omega}\,.
\end{equation}
%%%%%%%%%%%%%%%%%%%%%%%%%%%%%%%%%
%
If we require $\meff\simeq 0$, the factor $K$, 
which depends on the see-saw parameters $y$, $z$, $\omega$ and $M_{1}$,
is expressed only in terms of the neutrino 
oscillation parameters:
%%%%%%%%%%%%%%%%%%%%%%%%%%%%%%%%%%
\begin{eqnarray}
K \simeq \frac{\cos2\theta_{12} + i\,\sin2\theta_{12}\cos\frac{\alpha_{21}}{2}}
{1 + \sin2\theta_{12}\sin\frac{\alpha_{21}}{2}}\,.
\label{meff0IH}
\end{eqnarray}
%%%%%%%%%%%%%%%%%%%%%%%%%%%%%%%%%%%%
%
If the Majorana phase $\alpha_{21}$ takes a CP-conserving value we get:
%%%%%%%%%%%%%%%%%%%%%%%%%%%%%%%%%%
\begin{eqnarray}\label{meffpiIHC}
K  &\simeq& \frac{\cos2\theta_{12} }{1+\eta_{k} \sin\left(2\theta_{12}\right)}\;\;\;\;\;{\rm for}\;\;\;\;\alpha_{21}=(2k+1)\pi\,,\;\;\;\;k=0,\pm1,\ldots\\
	K  &\simeq& e^{2i\eta_{k}\theta_{12}}\;\;\;\;\;{\rm for}\;\;\;\;\alpha_{21}=2k\pi\,,\;\;\;\;k=0,\pm1,\ldots\,,
\label{meff0IHCP}
\end{eqnarray}
%%%%%%%%%%%%%%%%%%%%%%%%%%%%%%%%%%%%
where $\eta_{k}= (-1)^k$.
In the case of  $\mefff^{A_1} = 0$ we obtain for  $|\mefff^{A_2}|$:
%%%%%%%%%%%%%%%%%%%%%%%%%%%%%%%%%%
\begin{equation}
 |\mefff^{A_2}| \cong \left |1 - \frac{f(A_2)}{f(A_1)}\right |\,
c^2_{13}\sqrt{|\Delta m^2_{\rm A}|}\,
\left (1 - 
\sin^22\theta_{12}\,\sin^2\frac{\alpha_{21}}{2}\right )^{\frac{1}{2}}\,.
\label{meffIHA2}
\end{equation}
%%%%%%%%%%%%%%%%
% 
 
 The condition (\ref{meff0IH}) (or (\ref{meffpiIHC}) and (\ref{meff0IHCP}))
for  $\mefff^{A} = 0$ strongly constrains the 
phase $\omega$ and the size of $K$. 
In order to be satisfied, the condition of cancellation
between $\mefff^{{\rm std}}$ and $\mefff^{{\rm N,A}}$
requires a correlation between the values 
of the see-saw parameters 
$y$, $z$, the phase $\omega$ and $M_{1}$, and the 
neutrino mass $\sqrt{|\dma|} = m_2$, 
the solar neutrino mixing angle
$\theta_{12}$ and the Majorana phase $\alpha_{21}$.
A priori, such a correlation seems highly unlikely,
making the cancellation in the effective Majorana 
mass in the scheme considered appear unprobable. 
Thus, if the light neutrino mass 
spectrum is with inverted hierarchy, within the 
see-saw scheme considered with two heavy Majorana 
neutrinos, it appears quite unlikely that 
if, for example, the GERDA III experiment with  $^{76}$Ge
observes a positive $\betabeta$-decay signal,  
another $\betabeta$-decay experiment
which uses a nucleus different from $^{76}$Ge, 
will not see a signal due to a strong suppression 
of the effective Majorana mass 
caused by the cancellation under 
discussion for that nucleus,
and vice-versa.

\begin{figure}
\begin{center}
\begin{tabular}{c}
\begin{tabular}{cc}
\includegraphics[width=7.5cm,height=6.5cm]{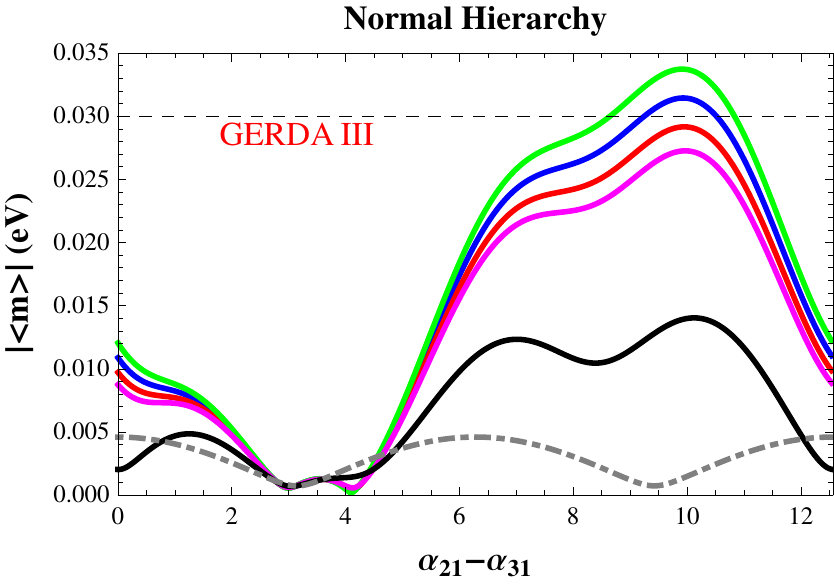} &
\includegraphics[width=7.5cm,height=6.5cm]{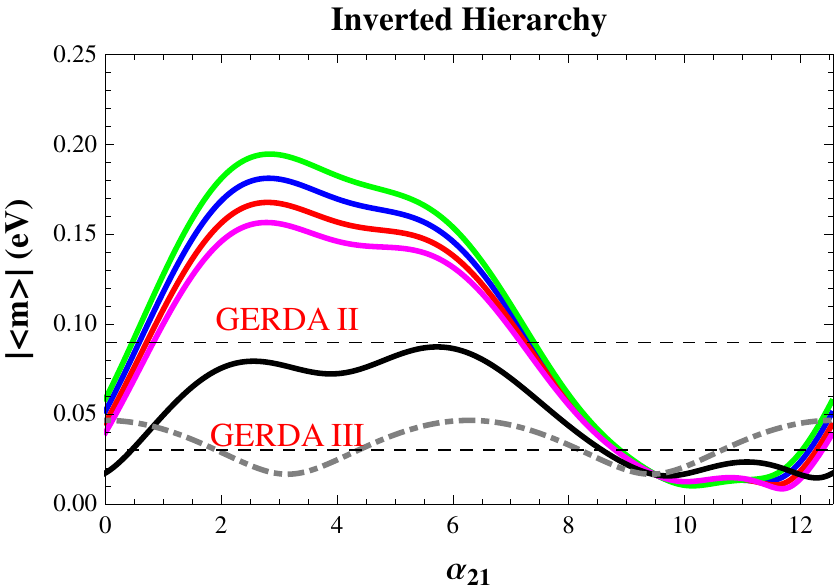}\\
\includegraphics[width=7.5cm,height=6.5cm]{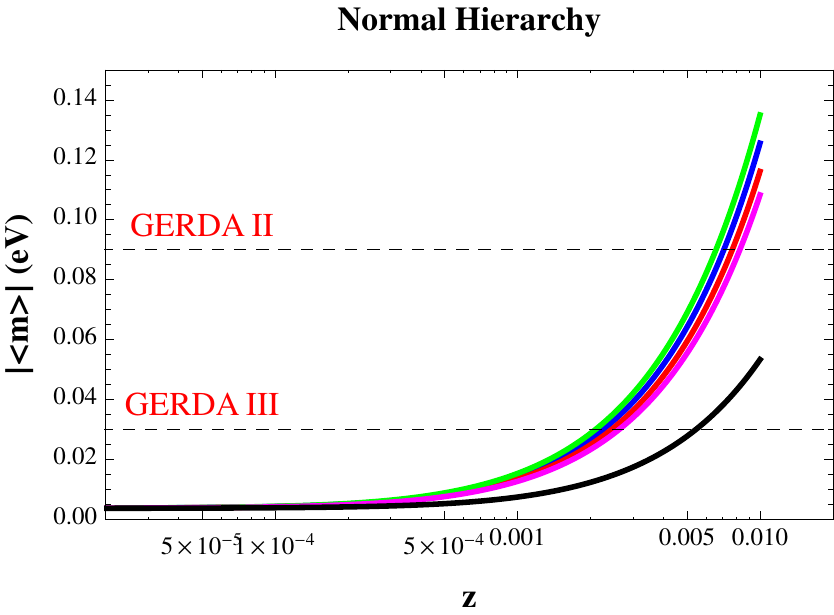} &
\includegraphics[width=7.5cm,height=6.5cm]{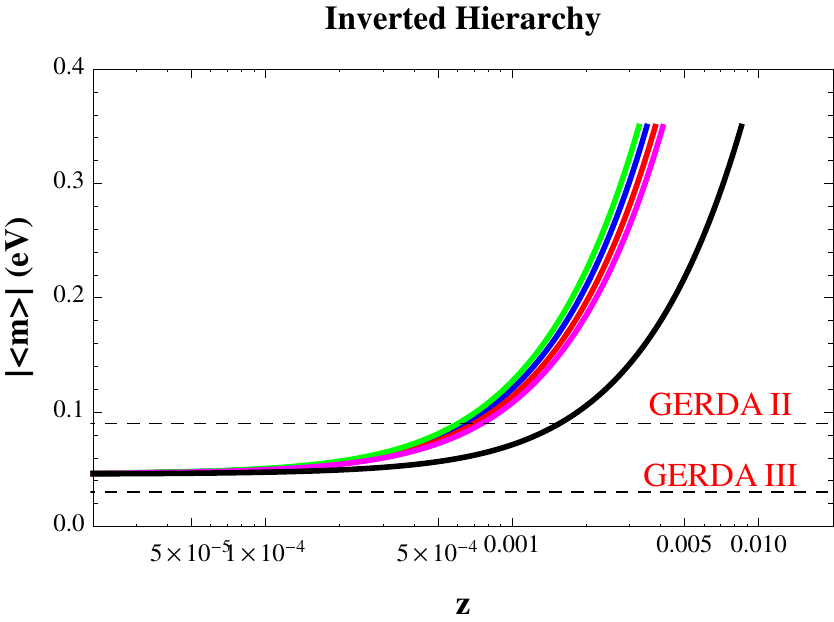}
\end{tabular}\\
\includegraphics[width=6cm,height=2cm]{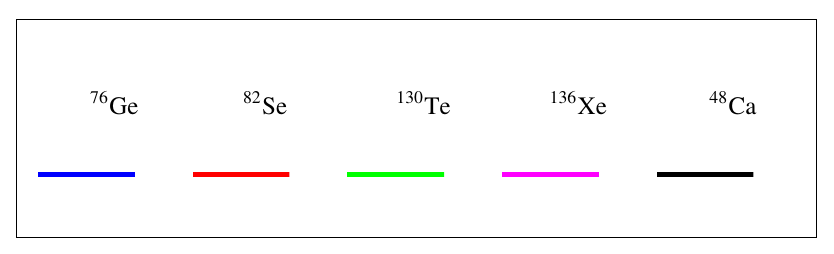}
\end{tabular}
\caption{Upper panels: the dependence of the 
effective Majorana mass 
$\meff$ on the Majorana phase $\alpha_{21}-\alpha_{31}$ (left side),
$\alpha_{21}$ (right side)
for  $M_{1}=100$ GeV, $\omega=\pi/4$, $y=0.01$ and $z=10^{-3}$.  The gray 
dash-dotted line shows $|\mefff^{\rm std}|$.
Lower panels: 
$\meff$ versus the degeneracy parameter $z$ 
for $M_{1}=100$ GeV, $\omega=0$, $y=0.01$ and $\alpha_{21}=0$. 
The left (right) panels correspond to a NH (IH) 
light neutrino mass spectrum. 
\label{fig5}}
\end{center}
\end{figure}

  The dependence of the interplay between 
 $\mefff^{{\rm std}}$ and $\mefff^{{\rm N}}$,
$i.e.$ between the contributions to 
$\meff$ due to the exchange of light and 
heavy Majorana neutrinos,
on  $z$, $\omega$ and the effective Majorana 
phase $\alpha_{21}$ (IH spectrum) or $\alpha_{21}-\alpha_{31}$
(NH spectrum)
is illustrated in Fig.~\ref{fig5} for $M_{1}=100$ GeV
and $y=0.01$. The solar and atmospheric neutrino oscillation 
parameters, $i.e.$ ($\theta_{12}$, $\dmsol$) and 
($\theta_{23}$, $\dma$), respectively, 
are fixed to their corresponding best fit values, 
and we have set $\theta_{13}=0.2$ and $\delta=0$.
Notice that the plots showing the correlation 
of $\meff$ and the Majorana phase
are symmetric with respect to 
$\alpha_{21}$ or $(\alpha_{21}-\alpha_{31})$ equal 
to $\pi/2$ and $3\pi/2$ if the phase $\omega$ 
takes the values $\omega=0,\pi/2,\pi$.
As Fig.~\ref{fig5} shows, we can have 
$\meff \gtap 0.01$ eV in the case of NH 
spectrum for a relatively large 
range of values the Majorana phase 
$\alpha_{21}-\alpha_{31}$ 
and certain values of the phase 
$\omega$: for, e.g.,  $\omega =\pi/4$, 
we get $\meff \gtap 0.01$ eV
for  $5\ltap \alpha_{21}-\alpha_{31}\ltap 4\pi$ 
if $z = 10^{-3}$. 
Actually, for the indicated 
values of $z$ and $\alpha_{21}-\alpha_{31}$ 
we have $\meff\gtap 0.01$ eV 
for any value of $\omega$ 
from the interval $[0,2\pi]$.
The interplay between the Majorana 
phase and $\omega$ can induce also a minimum
of $\meff$ for a certain value 
of the degeneracy parameter $z$.
For example, in the case of the 
IH spectrum, for $\alpha_{21}=0$ and $\omega=\pi/3$,
the predicted effective Majorana mass can be smaller than
$0.03$ eV: we have for the different nuclei considered
$0.007\,{\rm eV}\lesssim\meff\lesssim 0.03$ eV if
$2\times 10^{-4}\lesssim z \lesssim 10^{-3}$.

%%%%%%%%%%%%%%%%%%%%%%%%%%%%%%%
%%%%%%%%%%%%%%%%%%%%%%%%%%%%%%%%%%%%

%%%%%%%%%%%%%%%%%%%%%%%%%%%%%%%
%%%%%%%%%%%%%%%%%%%%%%%%%%%%%%%%%%%%

%%%%%%%%%%%%%%%%%%%%%%%%%%%%%%%
%%%%%%%%%%%%%%%%%%%%%%%%%%%%%%%%%%%%

\mathversion{bold}
\section{The $\betabeta$-Decay Effective Majorana Mass and
$B(\mu\to e+\gamma)$}
\mathversion{normal}

We now combine the information on the see-saw 
parameter space that we obtained from the 
lepton flavour ($\mu\to e+\gamma$) and lepton 
number ($\betabeta$) violating processes 
studied in the previous sections.
 In the simple extension of the Standard Model 
considered so far, $i.e.$ with the addition of two 
heavy RH neutrinos $N_{1}$ and $N_{2}$ at the TeV scale, 
which behave as a pseudo-Dirac particle and at 
the same time are responsible for the
generation of neutrino masses via the see-saw mechanism, 
a sizable (dominant) contribution of $N_{1}$ and $N_{2}$ 
to the $\betabeta$-decay rate would  imply a ``large'' 
effect in the muon radiative decay rate. 
 Indeed, if $\meff \cong |\mefff^{{\rm N}}|$,
where $\mefff^{{\rm N}}$ is given in eq. (\ref{mee3}),
using eqs.~(\ref{Bmutoeg1}) and (\ref{T3b}) 
it is easy to show that, 
given the splitting 
$10^{-4}\lesssim z\ll 1$ between $M_1$ and $M_2$,
$\meff \cong |\mefff^{{\rm N}}|$ can be directly 
related to the $\mu \to e+\gamma$ decay branching ratio.
More explicitly, we have:
%%%%%%%%%%%%%%%%%%%%%%%%%%%%%%%%%
\begin{equation}
	B(\mu\to e+\gamma)\;\cong\;
	\frac{3\alpha_{\rm em}}{64\pi}\,\left| G(0)-G(X)\right|^{2}\,\left| 
r \right|^{2}\, \frac{M_{1}^{2}}{M_{a}^{4}}\,
\frac{|\mefff^{{\rm N}}|^2}{z^{2} (f(A))^{2}}\,,	
\label{Bmeg}
\end{equation}
%%%%%%%%%%%%%%%%%%%%%%%%%%%%%%%%
%
where 
$r\equiv (U_{\mu 2}-i\sqrt{m_{3}/m_{2}}\,U_{\mu 3})/
(U_{e 2}-i\sqrt{m_{3}/m_{2}}\,U_{e 3})$ 
for the NH mass spectrum.
As was pointed out earlier, the corresponding expression 
for the case of IH spectrum is obtained by replacing
$m_{2,3}\to m_{1,2}$ and  
$U_{\alpha 2, \alpha 3}\to U_{\alpha 1,\alpha 2}$ ($\alpha=e,\mu$). 
For the NH (IH) neutrino mass spectrum one 
has: $0.5 \lesssim|r|\lesssim 30$ ($0.01 \lesssim|r|\lesssim 5$).
Therefore, for $M_{1}\approx 100$ GeV and NH spectrum,
the predicted rate of the $\mu \rightarrow e + \gamma$
decay can even be larger by up to one order of magnitude 
than the upper bound on $B(\mu\to e+\gamma)$ if
a positive signal in the current experiments searching 
for  $\betabeta$-decay is detected, implying 
$\meff\sim 0.1$ eV.

  A lower bound on $B(\mu\to e+\gamma)$ in (\ref{Bmeg}) 
can be derived for both types of light 
neutrino mass spectrum.
In the case of NH spectrum, such lower bound 
is within the sensitivity of the MEG experiment, 
 provided a $\betabeta$-decay corresponding to 
$\meff\gtap 2\times 10^{-2}$ eV is observed. 
%%%%%%%%%%%%%%%%%%%%%%%%%%%%%%%%%
\begin{figure}
\begin{center}
\includegraphics[width=13.5cm,height=9.5cm]{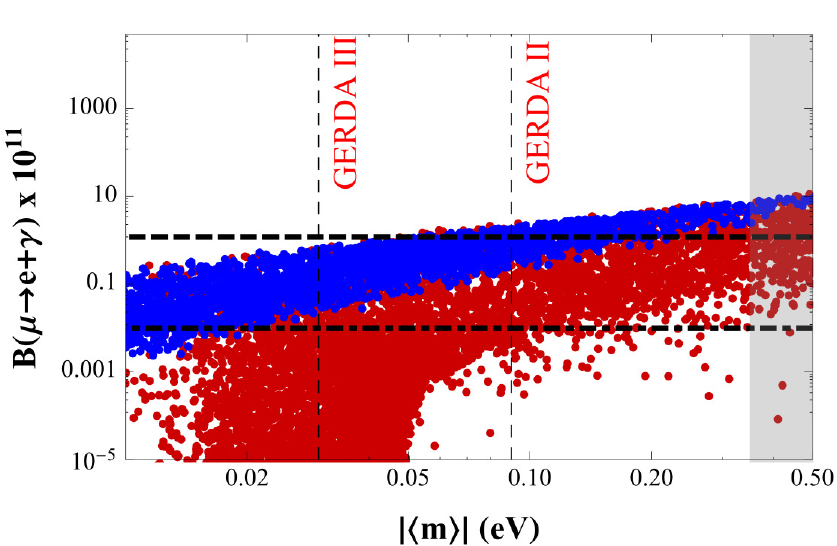}
\caption{$B(\mu\to e+\gamma)$ vs $\meff$ for $M_{1}=100$ GeV,  
$z=10^{-3}$ and $i)$ NH neutrino mass spectrum 
(blue dots), $ii)$ IH neutrino mass spectrum (red dots).\label{fig6}}
\end{center}
\end{figure}
%%%%%%%%%%%%%%%%%%%%%%%%%%%%%%%%%%%%%%
%

The analytic relation between $B(\mu\to e+\gamma)$
and  $\meff$  in eq. (\ref{Bmeg}) 
is confirmed by the results of numerical 
computation, reported in Fig.~\ref{fig6}.
The plot shows the correlation between the $\mu\to e+\gamma$
branching ratio and the effective Majorana mass in the 
case of ``large'' couplings between 
the RH (pseudo-Dirac pair) neutrinos and 
charged leptons in the Lagrangian (\ref{NCC}).
 The effective Majorana mass $\meff$ was computed 
for $z=10^{-3}$ using the general expression (\ref{mee1}). 
The $\betabeta$-decay nucleus was assumed to be $^{76}$Ge. 
The neutrino oscillation parameters are taken, again, within the corresponding
$3\sigma$ experimental intervals reported in Table 1.
The Majorana phase $\alpha_{21}$ ($\alpha_{31} - \alpha_{21}$) 
and the phase $\omega$ in the IH (NH) case 
were varied in the intervals  $[0,4\pi]$ and $[0,2\pi]$, respectively. 
The neutrino Yukawa coupling takes values $y\lesssim 0.1$.  
The correlation between $B(\mu\to e+\gamma)$
and $\meff\cong |\mefff^{{\rm N}}|$ reported in eq.~(\ref{Bmeg})
is satisfied for values $y\gtrsim0.01$. This is in agreement with
Figs.~\ref{fig3} and \ref{fig4}, where it is shown
that a signal compatible with the GERDA sensitivity reach is 
possible, provided $y\gtrsim 10^{-3}$, for both types of neutrino mass
spectrum. 
Moreover, in the case of IH light neutrino mass spectrum, such correlation
depends strongly on the value of the Majorana phase
$\alpha_{21}$. Indeed, for $M_{1}\cong 100~(1000)$ GeV and
$y\cong 0.01~(0.1)$ we expect that the MEG experiment \cite{MEG} is able to measure
the $\mu\to e+\gamma$ decay rate (see Fig.~\ref{fig2}). If lepton flavour violation is 
discovered by MEG, according to eqs.~(\ref{mee3}) and (\ref{Ue21IH}),
a positive signal detected by GERDA II, $i.e.$ $\meff\cong|\mefff^{N}|\gtrsim 0.1$ eV,
implies: $10^{-3}~(10^{-2})\lesssim z(1+0.94\sin(\alpha_{21}/2))\lesssim 4\times 10^{-3}~(4\times 10^{-2})$.
In the case of $M_{1}=100$ GeV and $z=10^{-3}$, used to obtain  Fig.~\ref{fig6}, we would
expect, in general, positive signals to be observed in both MEG and GERDA II 
experiments if $\alpha_{21}\cong 0,\pi$;
in the case of  $\alpha_{21}\cong3\pi$,
the $\betabeta$  and $\mu \to e+\gamma$ decays are predicted to
proceed with rates below the sensitivity of these two experiments.

We note, however, that it is not possible to 
get independent constraints on the 
degeneracy parameter $z$ and the Majorana phase
from the data on $\betabeta$  and $\mu \to e+\gamma$ decays.
Finally, we notice also that the strong correlation exhibited 
in Fig.~\ref{fig6} is a consequence of the constraints 
imposed by the neutrino oscillation data on the  
type I see-saw parameter space in the case 
investigated by us.

\section{Conclusions}

We have analyzed the low energy implications 
of a type I see-saw scenario with right-handed (RH) neutrino masses at
the electroweak scale and sizable charged and neutral current
weak interactions. This class of scenarios have the attractive feature
that the RH neutrinos could be directly produced at 
the Large Hadron Collider, thus allowing to test in collider
experiments the mechanism of neutrino mass generation. Furthermore,
and in contrast to the high-scale see-saw mechanism, the rates for 
the rare leptonic decays are unsuppressed in this scenario, which
opens up the possibility of detecting signatures of new physics with
experiments at the intensity frontier.

Present low energy data set very stringent constraints on this scenario. Namely, 
reproducing the small neutrino masses requires, barring cancellations, 
that two of the heavy RH neutrinos must form a pseudo-Dirac pair even in the 
case when there is no conserved lepton charge
in the limit of zero splitting at tree level
between the masses of the pair.
Besides, reproducing the experimentally
determined values of the low energy neutrino oscillation parameters (mixing angles and neutrino mass squared differences)
fixes the weak charged current (CC) and neutral current (NC) couplings of the heavy Majorana neutrinos to the charged leptons and $W^{\pm}$ and light neutrinos and $Z^{0}$, $(RV)_{\ell k}$ (see eqs. (\ref{NCC}) and (\ref{NNC})),
up to an overall scale which can be related to the largest eigenvalue $y$ of the matrix of neutrino Yukawa couplings.
This allows to derive explicit expressions for the rates of the lepton
flavour violating (LFV) charged lepton radiative decays,  
$\mu \to e + \gamma$, $\tau \to e + \gamma$, $\tau \to \mu + \gamma$,
in terms of the low energy (in principle measurable) neutrino mixing parameters (including the Dirac and
Majorana CP violating phases), the neutrino Yukawa coupling $y$ and the RH neutrino mass scale. 
Using the present constraint on the rate of the process $\mu \to e + \gamma$ we have obtained an
upper bound on the coupling $y$ under the assumption that the RH neutrino
mass scale is in the range ($100\div 1000$) GeV. Our analysis shows that the restrictions on
this scenario from the data on the neutrino mixing parameters and the upper bound on 
the $\mu \to e + \gamma$ decay rate imply that the CC and NC
couplings $|(RV)_{\ell k}|$ of the heavy RH Majorana neutrinos are too small to allow their production 
at the LHC with an observable rate.
Other lepton flavour violating processes, such as $\mu-e$ conversion in nuclei, give complementary 
constraints on the parameter space of this model, which will be discussed elsewhere \cite{mu2e}.

We have also analyzed the enhancement of the rate of neutrinoless 
double beta ($\betabeta$-) decay induced by the RH neutrinos.
We have  shown that even after imposing the restrictions 
on the parameter space implied by the 
data on the neutrino oscillation parameters 
and on the LFV charged  lepton radiative decays, 
the contribution due to the 
exchange of the RH neutrinos 
in the $\betabeta$-decay amplitude 
can substantially enhance the 
$\betabeta$-decay rate. As a consequence, the latter
can be in the range of 
sensitivity of the GERDA experiment
even when the light neutrinos possess a 
normal hierarchical mass spectrum. 
Finally, the rate of the 
$\mu\to e+ \gamma$ decay, generated by the
exchange of the heavy RH neutrinos can naturally be 
within the sensitivity of the MEG experiment.
Thus, the observation of $\betabeta$-decay in the 
next generation of experiments which are under 
preparation at present, 
and of the $\mu\to e+ \gamma$ decay 
in the MEG experiment, could be the 
first indirect evidence for the 
TeV scale type I see-saw mechanism of neutrino 
mass generation.

\begin{acknowledgments}
This work was supported in part by the INFN program 
on ``Astroparticle Physics'', by the Italian MIUR program 
on ``Neutrinos, Dark Matter and  Dark Energy in the Era of LHC'', 
by the World Premier International 
Research Center Initiative (WPI Initiative), 
MEXT, Japan  (S.T.P.) and by the DFG cluster of 
excellence “Origin and Structure of the Universe” (A.I.).
The work of E.M. is supported by
Funda\c{c}\~{a}o para a Ci\^{e}ncia e a
Tecnologia (FCT, Portugal) through the projects
PTDC/FIS/098188/2008,  CERN/FP/116328/2010
and CFTP-FCT Unit 777,
which are partially funded through POCTI (FEDER).
\end{acknowledgments}

\end{document}